\documentclass[twocolumn]{aastex631}

\newcommand{\kms}{\,km\,s$^{-1}$} 

\begin{document}

\title{The MUSE Ultra Deep Field (MUDF). VI. The relationship between galaxy properties and metals in the circumgalactic medium}

\correspondingauthor{Alexander Beckett}
\email{abeckett@stsci.edu}

\author[0000-0001-7396-3578]{Alexander Beckett}
\affiliation{Space Telescope Science Institute,
3700 San Martin Drive,
Baltimore, MD, 21218 USA}

\author[0000-0002-9946-4731]{Marc Rafelski}
\affiliation{Space Telescope Science Institute, 
3700 San Martin Drive, 
Baltimore, MD, 21218 USA}
\affiliation{Department of Physics and Astronomy, Johns Hopkins University, Baltimore, MD 21218,USA}

\author[0000-0002-4917-7873]{Mitchell Revalski}
\affiliation{Space Telescope Science Institute, 
3700 San Martin Drive, 
Baltimore, MD, 21218 USA}

\author[0000-0001-6676-3842]{Michele Fumagalli}
\affiliation{Dipartimento di Fisica G. Occhialini, Universit\`a degli Studi di Milano-Bicocca, Piazza della Scienza 3, I-20126 Milano, Italy}
\affiliation{INAF - Osservatorio Astronomico di Trieste, via G. B. Tiepolo 11, I-34143 Trieste, Italy}

\author[0000-0002-9043-8764]{Matteo Fossati}
\affiliation{Dipartimento di Fisica G. Occhialini, Universit\`a degli Studi di Milano-Bicocca, Piazza della Scienza 3, I-20126 Milano, Italy}
\affiliation{INAF - Osservatorio Astronomico di Brera, via Bianchi 46, I-23087 Merate (LC), Italy}

\author[0000-0001-5294-8002]{Kalina Nedkova}
\affiliation{Space Telescope Science Institute, 
3700 San Martin Drive, 
Baltimore, MD, 21218 USA}
\affiliation{Department of Physics and Astronomy, Johns Hopkins University, Baltimore, MD 21218,USA}

\author[0000-0002-6095-7627]{Rajeshwari Dutta}
\affiliation{IUCAA, Postbag 4, Ganeshkind, Pune 411007, India}


\author[0000-0001-9070-9969]{Rich Bielby}
\affiliation{Department for Education, Bishopsgate House, Feethams, Darlington DL1 5QE, UK}

\author[0000-0001-5804-1428]{Sebastiano Cantalupo}
\affiliation{Dipartimento di Fisica G. Occhialini, Universit\`a degli Studi di Milano-Bicocca, Piazza della Scienza 3, I-20126 Milano, Italy}

\author[0000-0001-8460-1564]{Pratika Dayal}
\affiliation{Kapteyn Astronomical Institute, University of Groningen, P.O. Box 800, 9700 AV Groningen, The Netherlands}

\author[0000-0003-3693-3091]{Valentina D'Odorico}
\affiliation{INAF - Osservatorio Astronomico di Trieste, via G. B. Tiepolo 11, I-34143 Trieste, Italy}
\affiliation{Scuola Normale Superiore, Piazza dei Cavalieri 7, I-56126, Pisa, Italy}

\author[0000-0002-1843-1699]{Marta Galbiati}
\affiliation{Dipartimento di Fisica G. Occhialini, Universit\`a degli Studi di Milano-Bicocca, Piazza della Scienza 3, I-20126 Milano, Italy}

\author[0000-0002-4288-599X]{C\'eline P\'{e}roux}
\affiliation{European Southern Observatory, Karl-Schwarzschild-Str. 2, 85748 Garching-bei-M\"unchen, Germany}
\affiliation{Aix Marseille Universit\'e, CNRS, LAM (Laboratoire d'Astrophysique de Marseille) UMR 7326, 13388, Marseille, France}




\begin{abstract}

We present intial results associating galaxies in the MUSE Ultra Deep Field (MUDF) with gas seen in absorption along the line-of-sight to two bright quasars in this field, to explore the dependence of metals in the circumgalactic medium (CGM) on galaxy properties. The MUDF includes $\sim$140h of VLT/MUSE data and 90 orbits of HST/G141M grism observations alongside VLT/UVES spectroscopy of the two quasars and several bands of HST imaging. We compare the metal absorption around galaxies in this field as a function of impact parameter, azimuthal angle and galaxy metallicity across redshifts 0.5 $<$ z $<$ 3.2. Due to the depth of our data and a large field-of-view, our analysis extends to low stellar masses ($<$ $10^{7}$ M$_{\odot}$) and high impact parameters ($>$ 600 kpc). We find a correlation between absorber equivalent width and number of nearby galaxies, but do not detect a significant anti-correlation with impact parameter. Our full sample does not show any significant change in absorber incidence as a function of azimuthal angle. However, we do find a bimodality in the azimuthal angle distribution of absorption at small impact parameters ($<$2 r$_{vir}$) and around highly-star-forming galaxies, possibly indicating disk-like accretion and biconical outflows. Finally, we do not detect any systematic deviation from the fundamental metallicity relation (FMR) among galaxies with detected absorption. 
This work is limited by gaps in the wavelength coverage of our current data; broader-wavelength observations with JWST will allow us to unlock the full potential of the MUDF for studying the CGM.

\end{abstract}



\section{Introduction} \label{sec:intro}

The exchange of gas between galaxies and their environment plays a vital role in galaxy assembly and evolution. Gas accretes onto galaxies from the surrounding circumgalactic and intergalactic medium (CGM and IGM), providing fuel for star formation, whilst supernovae and active galactic nuclei (AGN) shock gas in the interstellar medium (ISM) and drive it out of the galaxy \citep[e.g.][]{tumlinson2017}. This ejection of gas from the ISM is found to play a role in the `quenching' of star formation in galaxies \citep[e.g.][]{roberts-borsani2020, davies2021a, garling2024}, although interactions between galaxies such as ram-pressure stripping can also lead to quenching \citep[e.g.][]{wetzel2013, fossati2019a, kuschel2022}.

This picture of gas recycling is supported by several lines of evidence, including 'down-the-barrel' absorption measurements indicating outflows or accretion \citep[e.g.][]{steidel2010, rubin2012, rubin2014, xu2023}, absorption in multiple lines-of-sight around the same galaxy \citep[either using multiple background sources or lensed images e.g.][]{keeney2013, lopez2018, lopez2020, mortensen2021}, and direct CGM emission measurements using integral field units such as MUSE and KCWI \citep[e.g.][]{burchett2021, zabl2021, chen2021a, zhang2023, dutta2023a, guo2023}. Other studies of large galaxy--absorber samples find increased incidence/covering fraction of absorption close to galaxy major and minor axes \citep[e.g.][]{bordoloi2011, bouche2012, kacprzak2012, schroetter2019}. These are expected to trace accreting and outflowing gas respectively, as evidenced by the co-rotation between galaxies and absorbers near the major axis \citep[e.g.][]{ho2017, martin2019, french2020}, and increased incidence of metal absorption tracing warm ionized gas (such as O~VI 1032/1037 \AA) near the minor axis \citep[e.g.][]{ng2019}. These results at z $\lesssim$ 1.5 are reproduced by simulations, with most major hydrodynamical simulations finding major-axis inflows and minor-axis outflows on $\sim$100 kpc scales \citep[e.g.][]{defelippis2020, peroux2020a, hopkins2021}, and often further from the central galaxy.

However, these dependencies on azimuthal angle are most often found in high-column-density gas (high-equivalent-width absorbers) in close proximity to galaxies. Studies focusing on lower-column-density absorption at larger scales have not detected any difference in absorber equivalent width at different azimuthal angles \citep[e.g][]{dutta2020, huang2021a}. \citet{pointon2019} found only a tentative azimuthal dependence in absorber detection for the galaxy--sightline pairs with metal detection, and no dependence for those without, again at low redshifts. There is also some evidence that at redshifts z $\gtrsim$ 2, the gas distribution is more isotropic, with outflows less collimated along galaxy minor axes \citep[e.g.][]{chen2021a}.

If outflowing and infalling material are spatially separated and recycle over long timescales, we would expect to see a higher metallicity in outflowing material. Some studies do show evidence for this \citep[e.g.][]{wendt2021}, yet others have found that neither the metallicity nor ionization state of absorbing gas depends significantly on azimuthal angle \citep[e.g.][]{kacprzak2019, pointon2019, weng2023a}, perhaps due to integrating multiple `clouds' along the line-of-sight \citep[e.g.][]{ramesh2024}.

Overall, at low redshifts gas recycling does appear to manifest as biconical outflows and co-rotating accretion around galaxies. The extent to which these structures can be found and how this extent varies with galaxy properties and environments remains unclear, as does the behaviour of outflowing and accreting material at higher redshifts. This relative lack of constraints from observations results in simulations with differing CGM properties, such as mass outflow rates from the CGM to the IGM exceeding those from the ISM to the CGM in both the EAGLE and FIRE-2 simulations \citep{mitchell2020, pandya2021}, whilst studies based on the TNG50 simulations find decreasing mass loading factor with increasing radius \citep{nelson2019}\footnote{Although mass loading at injection is part of the input to TNG, the outflows are resolved at impact parameters larger than $\sim$10kpc.}. This could be impacted by the changing cosmic star-formation rate \citep[e.g][]{madau2014}, as well as the influence of cosmic rays \citep[e.g.][]{rodriguezmontero2023}, and may result in outflows reaching Mpc scales.

In addition to forming structures in the CGM, this feedback also affects the metallicities of the ISM of galaxies. This manifests as the mass-metallicity relation \citep[MZR, e.g.][]{andrews2013, somerville2015, curti2020, sanders2021}, a positive correlation which demonstrates that more massive galaxies retain more of their metals, whereas lower-mass galaxies more efficiently eject metals into the CGM and IGM \citep[e.g.][]{tortora2022, yang2022}. When galaxy star-formation rates are also taken into account, this forms the fundamental metallicity relation (FMR), which does not evolve substantially with redshift at z $\lesssim$ 3, suggesting that it is driven primarily by supernova feedback \citep[e.g.][]{dayal2013, sanders2018, curti2024}.

In order to constrain the presence and properties of these CGM structures, and hence improve our understanding of stellar feedback and study its effects on galaxy evolution, observations of gas at large impact parameters are required over a range of redshifts.

\citet{beckett2021}, used a dataset covering a wide field around the Q0107 quasar triplet at redshifts z $\lesssim$ 1, finding a bimodality in the position angle distribution of H~I absorption alongside evidence from kinematics and metal content supporting disk and outflow models to scales of $\approx$ 300 kpc, with little evidence of such flows beyond that radius.

The MUSE Ultra Deep field, with deep spectroscopic observations around a pair of bright quasars at z$\approx$3.2, provides a unique opportunity to extend this analysis to higher redshifts, as well as low stellar masses down to $10^{6.5}$ M$_{\odot}$, and covers a variety of metal absorption lines.

In this paper we compare the properties of metal absorption in the CGM/IGM with galaxy properties. Section \ref{sec:data} summarizes the MUDF dataset (described fully across our series of papers: \citealt{lusso2019, fossati2019, lusso2023, revalski2023, revalski2024}; Fossati et al., in prep) followed by information on the resulting samples of galaxies and absorbers in Section \ref{sec:samples}. We then study the absorber distribution around galaxies as a function of azimuthal angle (Section \ref{sec:tests}), and match absorber properties with galaxy metallicity (Section \ref{sec:metallicity}). We discuss how these results can be interpreted in the context of other studies in Section \ref{sec:discussion}, before presenting some final remarks and conclusions in Section \ref{sec:conclusions}. 

Throughout this work we quote distances in physical units, and assume a $\Lambda$CDM cosmology with H$_{0}$ = 67.4 $\textrm{km} \, \textrm{s}^{-1} \, \textrm{Mpc}^{-1}$ and $\Omega_{m}$ = 0.315, as found by \citet{planckcollaboration2020}.

\section{Data} \label{sec:data}

\subsection{The MUSE Ultra Deep Field} \label{sec:mudf_summary}

An extremely deep galaxy survey is required in order to characterize the gas around galaxies with masses as low as $10^{6.5}$ M$_{\odot}$ and redshifts up to z $\approx$ 3. Therefore, in an effort to maximize the utility of such an expensive survey, we have chosen the field P2139-443,featuring two bright quasars (Q2139-4434, hereafter QSO-SE, and Q2139-4433, hereafter QSO-NW) at z $\approx$ 3.23. These have r-band magnitudes 17.9 and 20.5 respectively and are separated by only $\sim 1'$, corresponding to about 500 kpc at z $\approx$ 3. Two additional quasars are also present, one fainter QSO at a similar redshift (Q2138-4427, z $\approx$ 3.17), and a fourth at a much lower redshift (z $\approx$ 1.30). These additional quasars do not have high-resolution spectra so are not covered in this study. 

We use the Multi-Unit Spectroscopic Explorer \citep[MUSE,][]{bacon2010} on the Very Large Telescope (VLT) to conduct a blind galaxy survey around these quasars, alongside high-resolution spectroscopy of QSO-SE and QSO-NW using the Ultraviolet and Visual Echelle Spectrograph \citep[UVES,][]{dekker2000}. In addition to this optical data, we have obtained X-ray, NIR and sub-mm observations covering this field, including high-resolution imaging and deep grism spectroscopy using the Hubble Space Telescope (HST). These observations have been described in detail in previous works, 
and can be summarized as follows:

\begin{itemize}
    \item VLT/MUSE observations of the field totalling $\approx$ 142 hours on-sky, making it tied with the MUSE eXtremely Deep Field \citep[MXDF,][]{bacon2017, bacon2023} as the deepest field observed with MUSE to-date.  This is European Southern Observatory (ESO) program 1100.A-0528 (PI: Fumagalli), and provides observed-frame optical and NIR spectra in the wavelength range 4650-9300 \AA. 
    These data reach 3$\sigma$ line detections below 3 $\times$ 10$^{-19}$ ergs s$^{-1}$ cm$^{-2}$ in the deepest parts of the field. Whilst the observing strategy and data reduction are described in \citet{lusso2019} and \citet{fossati2019}, the final sources are presented in \citet{revalski2023}.

    \item VLT/UVES spectroscopy of the two bright quasars (QSO-SE and QSO-NW), covering the wavelength range 4100-9000 \AA{ } at high resolution (R $\sim$ 40,000) and detailed in \citet{fossati2019}. The total time on-sky is 7 hours for the brighter QSO-SE and $\approx$ 32 hours for the fainter QSO-NW, resulting in signal-to-noise per pixel of $\approx$ 25 and $\approx$ 10 respectively. These data were collected under ESO programs 65.O-0299, 68.A-0216, 69.A-0204, and 102.A-0194 (all PI: D'Odorico).

    \item HST imaging using multiple wide-band filters (F336W with WFC3/UVIS, F450W and F702W with WFPC2, and F125W and F140W with WFC3/IR) provides coverage from $\approx$ 3000 to 16000 \AA{ }(although with some gaps). 
    These images reach 5$\sigma$ detection levels between $m_{AB}$= 27 and 29. The data originate from proposals 6631 (PI: Francis), 15637 (PIs: Rafelski \& Fumagalli), and 15968 (PI: Fossati), and are described fully in \citet{revalski2023}. 

    \item HST grism spectroscopy using the G141 grism, approximately covering the wavelength range from 11,000 to 17,000 \AA{ }at a resolving power R $\approx$ 130. With $\approx$ 48 hours exposure time, continuum is usually detected at 5$\sigma$ significance down to $m_{F140W}$= 27. The redder coverage allows spectroscopic redshifts to be confirmed in parts of the MUSE `redshift desert' from 1.5 $<$ z $<$ 3. These also originate from proposal 15637 and are described in \citet{revalski2023}.

    \item VLT/HAWK-I (High Acuity Wide field K-band Imager, \citealt{pirard2004}) imaging of the field under ESO program 0105.A-0564 (PI: Fossati). This covers the field with $\approx$ 8 hours of data from 1.99 to 2.31 $\mu$m, reaching 5$\sigma$ depth of $m_{AB}$ $\approx$ 25.5. This is used in the SED-fitting routine from which we obtain stellar masses of galaxies in the MUDF, and will be described in Fossati et al. (in prep).

\end{itemize}

Data products arising from the MUSE data are available using the DOI:\dataset[10.18727/archive/84]{\doi{10.18727/archive/84}}. Products arising primarily from the HST data are available using: \dataset[10.17909/81fp-2g44]{\doi{10.17909/81fp-2g44}}.

This study makes use of the UVES data in order to study gas properties through absorption features observed in the quasar spectra, and then relates this to the nearby galaxies with properties determined from the MUSE, HST and HAWK-I data. The field has also been observed with XMM-Newton \citep{jansen2001} and the Atacama Large Millimeter/submillimeter Array (ALMA, \citealt{wootten2009}). These are used to help discriminate between star-forming galaxies and active galactic nuceli (AGN) as described in \citet{revalski2024}, and in results from \citet{lusso2023}.

We also use several derived properties from \citet{revalski2024} and Fossati et al. (in prep), which describe respectively galaxy ISM metallicities measured from a range of emission lines, and galaxy stellar masses and star-formation rates measured using stellar population synthesis (SPS) modeling. The derivation of gas and galaxy properties from the MUDF data is described below.

\subsection{Absorption Measurements} \label{sec:absorption}

Absorption lines were characterized using the procedure described in \citet[][their Section 6.1]{fossati2019}, which we summarize here. We utilize the ESPRESSO data analysis software \citep{cupani2016}, which fits a cubic spline to the continuum redward of Ly$\alpha$. We then visually identify absorption systems in the normalized spectra, first using the presence of doublets (most often Mg~II), and then other transitions commonly seen in quasar spectra, such as Mg~I and Fe~II. The wavelength range covering each absorption line system is extracted and used for fitting the absorption features.

We fit these regions of the spectrum using the Monte-Carlo Absorption Line Fitter (MC-ALF, \citealt{longobardi2023}). This fitting algorithm uses Bayesian statistics to determine the minimum number of Voigt profile components required to model the selected part of the observed spectrum. This models the spectrum using different numbers of components, then uses the Akaike Information Criterion \citep[AIC,][]{akaike1974} to evaluate the best fit. The model with the smallest number of components that has a likelihood ratio within 1:150 of that with the best (smallest) AIC is selected. The model allows for `filler' components to be included, accounting for blends with different absorption lines from absorption systems at different redshifts. We fit doublets such as Mg~II (2796/2803 \AA) and C~IV (1548/1550 \AA) simultaneously, as these should exhibit equal numbers of components.

This fit provides the equivalent width (EW, which can be used to calculate column density for unsaturated lines), redshift, and Doppler width of each transition, although we note that EW measurements for blended lines are often highly uncertain. After excluding regions within 3000 \kms{ }of each quasar to remove any proximity effect \citep[e.g.][]{wild2008}, this results in an absorber catalog consisting of 304 Voigt components across both quasar sightlines. These form 31 absorption systems of width usually $\lesssim$ 500 \kms{ }and separation $\gtrsim$ 3000 \kms, for which the total EW resulting from each absorption line can be more confidently measured. Our sensitivity limits vary only slightly with redshift; we are able to detect absorption at 3$\sigma$ signifcance down to EW limits of 0.03 and 0.003 \AA{ }in QSO-NW and QSO-SE respectively. Some examples of these fits are provided in \citet{fossati2019}.


\subsection{Galaxy Properties} \label{sec:galaxies}

The techniques used to measure the properties of galaxies in this field and produce the galaxy catalog used in this study are detailed in \citet{fossati2019} and \citet{revalski2023}, with further properties derived in \citet{revalski2024} and Fossati et al. (in prep). However, as these measurements are an essential part of this work, we also summarize them below.

\subsubsection{Redshifts}\label{sec:redshifts}

Galaxies were identified in the HST F140W image using the Source Extractor (\textsc{SExtractor}) software \citep{bertin1996}, as described in \citet{revalski2023}. The segmentation maps produced by \textsc{SExtractor} were then used to extract spectra from the MUSE cubes and the HST grism data. Spectroscopic redshifts were measured using the interactive fitting routine described in \citet{henry2021}, modified for this dataset as described in \citet{revalski2024}, and applied simultaneously to the MUSE and HST spectra.


Below z $\approx$ 1.5, all galaxies have MUSE coverage of the resolved [O~II] doublet or the [O~III] and H$\beta$ lines, and at 1.5 $\lesssim$ z $\lesssim$ 2.4 the [O~III] and H$\beta$ lines are covered by the grism spectra. We are therefore able to obtain unambiguous redshift measurements for emission-line galaxies at z $\lesssim$ 2.4. Identifications at redshifts 2.4 $\lesssim$ z $\lesssim$ 2.9 are based on a single strong line: unresolved [O~II] in the HST grism. These are accepted only where there is identification of H$\gamma$, or where there is clear detection of continuum flux in MUSE (ruling out Ly$\alpha$ as the origin of the strong line). For z $\gtrsim$ 2.9, we also have coverage of Ly$\alpha$ in MUSE, which has an asymmetric profile that allows confident identification. Thus our galaxy redshift measurements are all robust.

We note that Ly$\alpha$ has a systematic offset in wavelength due to radiative transfer effects \citep[e.g.][]{verhamme2018}, usually of a few hundred \kms. In several cases, another line (often [O~II]) is visible from which we can measure the systemic redshift, but there are 4 galaxies for which Ly$\alpha$ is the only identified emission line. However, none of these lie within 1000 \kms of our absorbers, so this additional redshift uncertainty does not affect our results.

\subsubsection{Morphologies}\label{sec:morphologies}

Galaxy morphologies were measured using \textsc{Statmorph} \citep{rodriguez-gomez2019}, an Astropy-affiliated code that uses improved PSF modeling to measure galaxy sizes and orientations more accurately than \textsc{SExtractor}. These measurements are detailed in \citet{revalski2023}. We use morphologies from the F140W image. This is the second-deepest of the HST filters, but is preferred over the deeper F336W filter as galaxies above z $\approx$ 3 drop out of F336W due to the Lyman break. Galaxies also often have different morphologies in the far-UV \citep[e.g.][]{lauger2005, vulcani2014, nedkova2021}, so we use the F140W to ensure we are probing the optical-NIR continuum.

We primarily use the \textsc{Statmorph} results to measure the position angle (PA) of the major axis of our sample galaxies, as well as their inclination. In order to confirm the accuracy of these methods, we also compare the galaxy PAs and inclinations with those measured using \textsc{Galfit} \citep{peng2002, peng2010a}. This comparison is described in Appendix \ref{sec:pa_measurements}. In most cases, \textsc{Statmorph} returns uncertainties $\lesssim 3^{\circ}$ in PA and $\lesssim 10^{\circ}$ in inclination, whereas the differences between \textsc{Statmorph} and \textsc{Galfit} suggest PA uncertainties $\lesssim 12^{\circ}$. Whilst both perform well in most cases, there are a small number of galaxies ($\approx$ 5\%) for which blended sources affect the results. We inspect each of these galaxies `by-eye' and select the more plausible inclination and azimuthal angle measurement.

\subsubsection{Stellar Masses and Star-Formation Rates} \label{sec:sps_models}

Stellar masses and star-formation rates are measured by fitting the galaxy spectra and photometry using stellar population synthesis (SPS) models. This process will be detailed in Fossati et al. (in prep), but is briefly described here.

We fit the broad-band HST and HAWK-I photometry simultaneously with the MUSE spectra where available, using the Monte Carlo Spectro-Photometric Fitter described in \citet{fossati2018}. This fits the spectra and photometry using \citet{bruzual2003} models with a \citet{chabrier2003} initial mass function, assuming an exponential star-formation history and \citet{calzetti2000} extinction law. These models include emission lines with ratios from \citet{byler2017}, and provide reliable estimates and uncertainties of galaxy stellar masses and star-formation rates.  

The galaxy templates are available with metallicities of solar, 40\% solar, and 20\% solar. Whilst many studies assume solar metallicity, for our range of galaxy masses and redshifts we attempt to determine the best template to use for each galaxy. We initially use the solar-metallicity template, but then adjust if neccessary. 

For galaxies with measured metallicities (see Section \ref{sec:metallicities}), we use the template with the closest metallicity to that measured value. For those without, we use the closest metallicity to the median found for the stack that corresponds to our initial stellar-mass measurement (as described in \citealt{revalski2024}), iterating again if fitting at the new metallicity produces a substantially different stellar mass.

Whilst the measurement uncertainties on the stellar masses and SFRs are both small, additional uncertainties arise from the choice of star-formation history and stellar physics models. We assume 0.2 dex uncertainty in the stellar mass estimate on each galaxy, which covers these model-dependent uncertainties (see e.g. \citealt{conroy2013}) and dominates over measurement uncertainties that are usually $<$0.05 dex in our sample. A comparison of star-formation rates estimated through direct calibrations \citep[e.g.][]{kennicutt1998, kewley2004} against those obtained from the modeling (which use both emission lines and UV flux where available) suggests uncertainties in our SFR measurements are typically $\approx$ 0.3 dex in most cases, which again are dominated by model uncertainties over measurement uncertainties of $<$0.05 dex. 

We use galaxy halo mass estimates only in order to calculate a virial radius. For this we use the stellar mass/halo mass relation from \citet{behroozi2010}, including an additional uncertainty of 0.25 dex.

\subsubsection{Metallicities}\label{sec:metallicities}

Given the depth and wavelength coverage of our galaxy survey, there are numerous sources for which we have strong detections of several emission lines, and can therefore calculate a metallicity. These measurements are detailed in \citet{revalski2024}.

There are several possible methods to calculate metallicity, using a variety of metal lines to estimate the abundances of different metals and different Hydrogen lines to estimate the Hydrogen content \citep[see e.g.][]{kewley2019}. 
We utilize the ratios between strong nebular lines that are visible in most sources. Specifically, we use a Bayesian method involving the [O~II] ($\lambda\lambda$ 3727, 3730 \AA), H$\beta$ ($\lambda$ 4862 \AA) and [O~III] ($\lambda\lambda$ 4959, 5007 \AA) emission lines, with the ratios between these lines calibrated using the results from \citet{curti2017}. \citet{revalski2024} validate these strong-line results using the 12 sources where the lines required for a `direct' method are observed, confirming that our metallicities are reliable. This includes oxygen lines at 1666 \AA{ }and 4363 \AA. Other methods involving [N~II]/H$\alpha$ are not usable due to the low resolution of the grism data. 

The strong oxygen lines are available in either the MUSE or HST spectroscopic data across a large fraction of the redshift range probed in this study, allowing us to estimate metallicities for 73 galaxies, of which 57 lie above the z $\approx$ 0.45 lower limit for detection of associated Mg~II absorption. These are mostly found to have metallcities 7.5 $<$ 12 + log(O/H) $<$ 8.7 (or between solar and 5\% solar metallicity).

\section{Galaxy and absorber samples}\label{sec:samples}

\subsection{Sample Construction}\label{sec:sample_construction}

\begin{table*}
\begin{center}
\caption{Summary of absorber properties. \label{table:absorbers}}
\begin{tabular}{c c c c c c}
 \hline Redshift & Line & Components & Total EW (\AA) & Other Ions & Galaxies \\\hline
 
QSO-SE & & & & & \\ 
0.6793 & Mg~II (2796) & 5 & 0.85 $\pm$ 0.03 & & 14 \\ 
0.8822 & Mg~II (2796) & 15 & 1.32 $\pm$ 0.01 & Mg~I & 6 \\
 & Fe~II (2382) & 2 & 0.147 $\pm$ 0.004 & & \\ 
0.9868 & Mg~II (2796) & 7 & 0.70 $\pm$ 0.01 & Mg~I & 2 \\
 & Fe~II (2382) & 4 & 0.117 $\pm$ 0.003 & & \\ 
1.0525 & Mg~II (2796) & 15 & 1.63 $\pm$ 0.03 & Mg~I & 14 \\
 & Fe~II (2382) & 11 & 0.360 $\pm$ 0.009 & &  \\
1.1539 & Mg~II (2796) & 7 & 1.14 $\pm$ 0.12 & Mg~I & 5 \\
& Fe~II (2382) & 3 & 0.32 $\pm$ 0.02 &  &  \\
1.5711 & Mg~II (2796) & 6 & 1.36 $\pm$ 0.19 & Mg~I, Zn~II, Cr~II & 4\\
& Fe~II (2382) & 7 & 0.80 $\pm$ 0.05 & & \\
1.7569 & Mg~II (2796) & 3 & 0.23 $\pm$ 0.01 & Al~II, Al~III & 3\\
& Fe~II (2382) & 1 & 0.019 $\pm$ 0.002 & & \\
2.1126 & Mg~II (2796) & 5 & 1.55 $\pm$ 0.04 & Mg~I, Mn~II, Cr~II, Zn~II, & 2 \\
& Fe~II (2382) & 3 & 0.78 $\pm$ 0.02 & Al~II, Al~III, Ni~II, Si~II & \\
2.2534 & Mg~II (2796) & 9 & 1.89 $\pm$ 0.06 & Mg~I, Al~II, Al~III, O~I,  & 8 \\
& Fe~II (2382) & 8 & 0.45 $\pm$ 0.01 & Si~II, C~II & \\ 
2.3797 & Fe~II (2382) & 3 & 0.148 $\pm$ 0.007 & Si~II, O~I, C~II, Al~II, Zn~II & 3 \\
2.3909 & C~IV (1548) & 4 & 0.091 $\pm$ 0.004 & & - \\
2.5459 & C~IV (1548) & 1 & 0.010 $\pm$ 0.003 & & -\\
2.6922 & C~IV (1548) & 4 & 0.076 $\pm$ 0.005 & & - \\
2.7342 & Fe~II (2382) & 2 & 0.107 $\pm$ 0.006 & Al~II, Si~II, Si~IV, O~I, C~II & - \\
& C~IV (1548) & 7 & 0.625 $\pm$ 0.018 & & - \\ 
2.7698 & C~IV (1548) & 1 & 0.083 $\pm$ 0.012 & & 1* \\
2.7929 & C~IV (1548) & 1 & 0.016 $\pm$ 0.001 & & -\\
2.8126 & C~IV (1548) & 2 & 0.034 $\pm$ 0.002 & & -\\
2.9139 & C~IV (1548) & 2 & 0.437 $\pm$ 0.017 & & -\\
3.0437 & C~IV (1548) & 7 & 0.149 $\pm$ 0.005 & & -\\
3.0816 & C~IV (1548) & 1 & 0.009 $\pm$ 0.001 & & -\\
3.1584 & C~IV (1548) & 2 & 0.021 $\pm$ 0.002 & & -\\ \hline
 QSO-NW & & & & & \\
0.8820 & Mg~II (2796) & 3 & 0.44 $\pm$ 0.01 & & 6 \\ 
1.8069 & Fe~II (2382) & 4 & 0.44 $\pm$ 0.01 & Mg~I, Mg~II & 3\\
2.2951 & Fe~II (2382) & 6 & 2.06 $\pm$ 0.04 & Al~II, Al~III, Mg~I, Mg~II & 3 \\
2.3554 & C~IV (1548) & 3 & 0.139 $\pm$ 0.006 & & 0\\
2.3680 & C~IV (1548) & 5 & 0.323 $\pm$ 0.012 & Si~II, C~II, Fe~II, Al~II & 0\\
2.4611 & C~IV (1548) & 1 & 0.056 $\pm$ 0.005 & & -\\
2.7620 & C~IV (1548) & 5 & 0.587 $\pm$ 0.017 & Si~IV & 5* \\
2.8518 & C~IV (1548) & 5 & 0.572 $\pm$ 0.014 & & -\\
3.0419 & C~IV (1548) & 4 & 0.400 $\pm$ 0.012 & Si~IV & - \\ \hline

\end{tabular}
\tablecomments{Absorber properties for the metals considered in this work. EWs are given only for the strongest transition of Mg~II, Fe~II and C~IV, as these are among the strongest lines and are present at all redshifts with detected absorption. Columns are: Redshift of the strongest absorption component; ion and rest-frame wavelength of transition (in \AA); number of detected components in the fit to this absorption system; total equivalent width of absorption components in the selected line; any other ions detected at the same redshift; number of detected galaxies within 500 \kms{ }(at any impact parameter). Note that the lack of wavelength coverage of strong emission lines above z $\approx$ 2.4 means that only the galaxies for which weaker lines could be detected are assigned a spectroscopic redshift. These are marked with a star. }
\end{center}
\end{table*}

\begin{figure*}
\includegraphics[trim={1.7cm 0.6cm 1.7cm 1.8cm},clip, width=\textwidth]{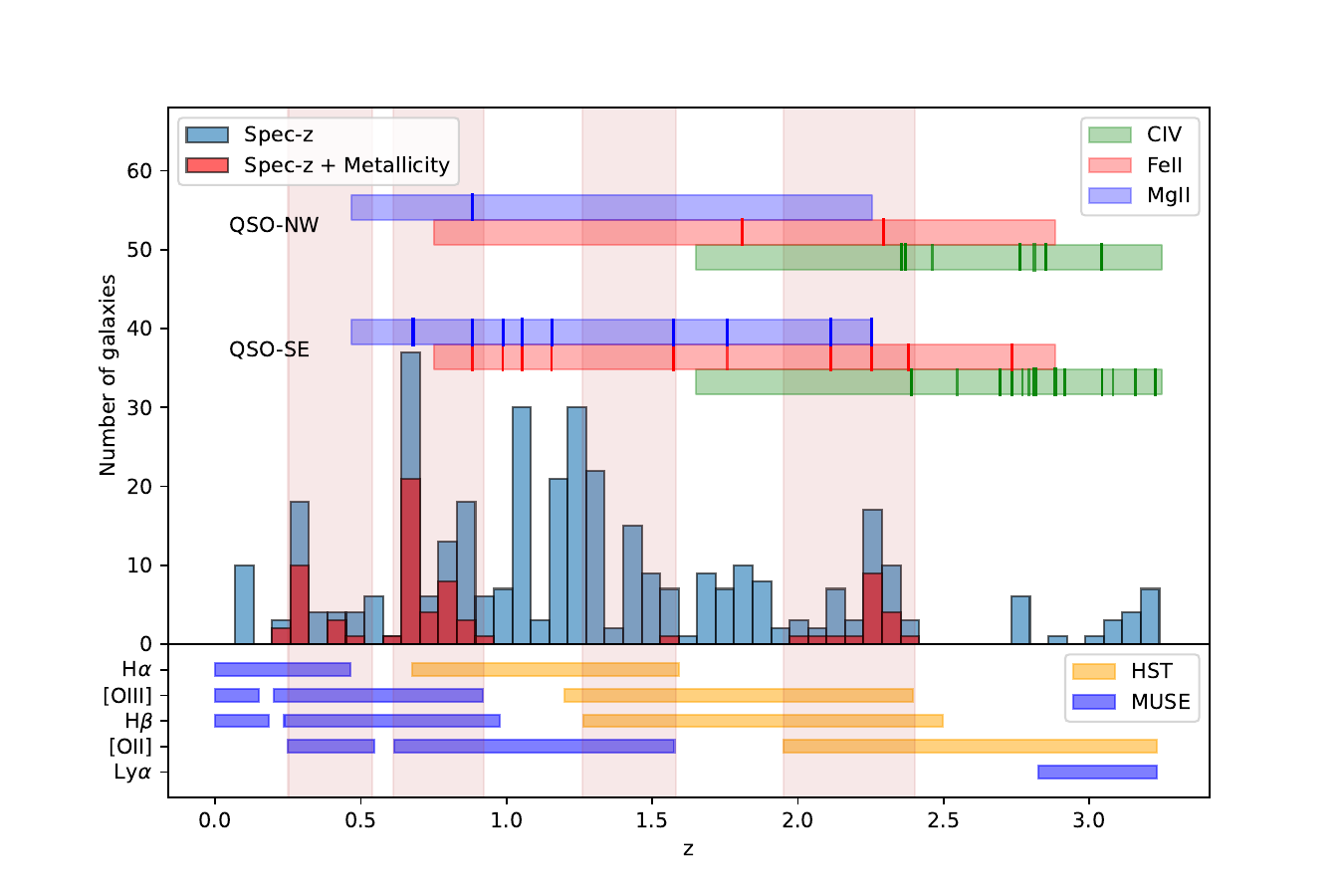}
\caption{The redshift distributions of galaxies and absorbers in the MUDF. Galaxies at or below the redshift of the quasar pair are shown; those with metallicity estimates are shown in red. The horizontal strips show UVES wavelength coverage of strong Mg~II, Fe~II and C~IV transitions in the QSO spectra, with the redshifts of detected absorption shown by the vertical lines. The width of these vertical lines is indicative of the total equivalent width of that ion at that redshift. The lower panel indicates the redshifts at which strong emission lines can be observed in MUSE (blue) and in the HST grism observations (yellow), and shows the gaps in wavelength coverage between MUSE and HST, as well as the MUSE AO laser gap. The red vertical bands indicate the redshifts at which the three lines required for metallicity measurements are accessible, and therefore the redshift ranges for which metallicities can be determined. 
\label{fig:z_dists}}
\end{figure*}

The above measurements result in a sample of 419 galaxies with spectroscopic redshifts and 304 Voigt components forming 31 absorption systems at different redshifts. Figure \ref{fig:z_dists} shows the redshift distribution of both samples. We note that all of our absorption systems feature at least one transition from Mg~II (usually the 2796/2803 \AA{ }doublet), Fe~II (2382 \AA) or C~IV (the 1548/1550 \AA{ }doublet), so for clarity only these ions are shown in the figure. Several other ions are detected in our absorption spectra, including Mg~I, Si~II-IV, Al II-III, Ni~II, Cr~II and Zn~II. The Mg~II, Fe~II and C~IV absorbers, along with any other detected ions, are listed in Table \ref{table:absorbers}. One absorber lies in close proximity to the quasar redshift, so is excluded from Table \ref{table:absorbers} and the rest of this work.

As shown in Figure \ref{fig:z_dists}, our wavelength coverage allows us to measure confident galaxy redshifts based on multiple emission lines from z=0 to the quasars, aside from 2.5 $\lesssim$ z $\lesssim$ 2.9. Unfortunately, this redshift range is also that in which many C~IV absorbers are found, limiting our ability to determine the galaxy environment around these absorbers. 

Galaxy metallicities can only be measured in the redshift ranges 0.3 $\lesssim$ z $\lesssim$ 0.95 (where [O~II], H$\beta$ and [O~III] are all visible in MUSE), 1.95 $\lesssim$ z $\lesssim$ 2.4 (where they are visible in the HST grism data), and a small range around z $\approx$ 1.5, where [O~II] lies in MUSE and H$\beta$ and [O~III] are visible in the HST grism.

\begin{figure*}
\includegraphics[trim={0.1cm 0.6cm 0.7cm 0.8cm},clip, width=\textwidth]{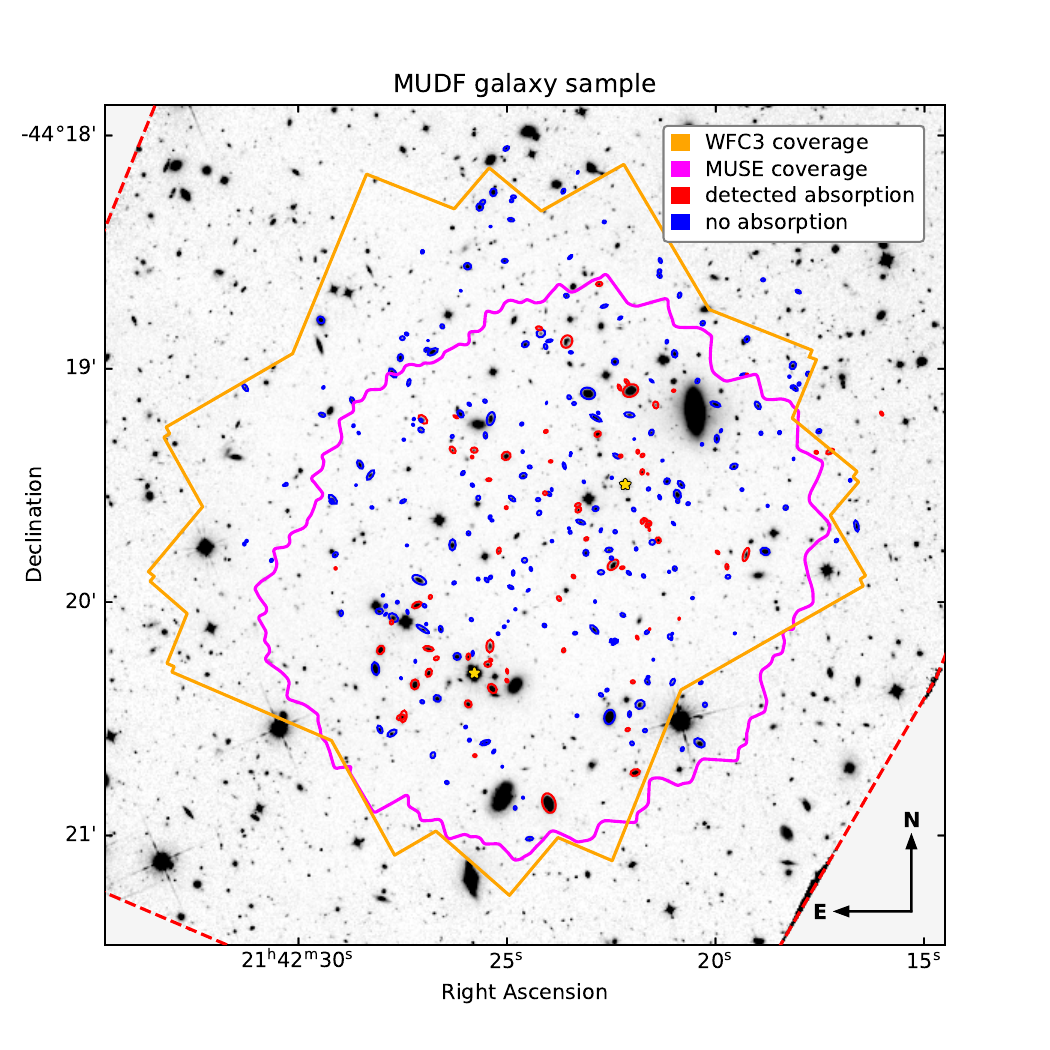}
\caption{Coverage of the MUDF using MUSE and WFC3 F140W, and galaxies with confirmed redshifts. The background image is from WFC3/F140W imaging, which has a shallow field (red outline) covering a much wider field than the grism and deep imaging (orange outline). The magenta line shows the limit of the MUSE coverage. Galaxies with spectroscopic redshifts 0.5 $<$ z $<$ 3.2 are also highlighted, as this redshift range is that at which absorption can be detected in our UVES spectra. Galaxies within 500 \kms of detected absorption are shown in red; those without detected absorption are in blue. The locations of the two quasars are indicated by yellow stars. We note that redshifts are more difficult to obtain using only the grism data, but the dispersion of the grism also allows redshifts to be determined for galaxies outside the field-of-view covered by the direct imaging. The MUSE observations cover a total area of 4.15 arcmin$^{2}$.
\label{fig:field_gals}}
\end{figure*}

Throughout this work, we associate absorbers with galaxies when the redshift difference between the galaxy and absorber is less than 500 \kms, although we generally confirm our results using a smaller 300 \kms{ }window. This matches the window used in \citet{beckett2021}, within which H~I absorption was found significantly more often around galaxies than in the field. Many other studies of absorbing gas use similar velocity cuts \citep[e.g.][]{bielby2019, prochaska2019, dutta2020, wilde2021, galbiati2023, karki2023}, although it may be conservative for low-mass galaxies with a low virial velocity. Some properties for those galaxies within 500 \kms{ }of detected absorption are provided in Table \ref{tab:gal_data}. Our galaxy sample is also shown in Figure \ref{fig:field_gals}, highlighting those galaxies within 500 \kms{ }of absorption. 

Many of our absorption systems consist of several different transitions at the same redshift. With different lines covered by our UVES spectra at different wavelengths, we must determine how best to select these in order to produce the fairest possible sample of galaxy--absorber pairs across a wide redshift range. At each redshift, we take the number of `absorbers' to be the number of identified Voigt components for whichever transition has the most components, producing a sample of 694 galaxy--absorber pairs. (This is usually Mg~II at lower redshifts and C~IV at higher redshifts.) This is a fairer test than counting the total number of components (which depends on the number of strong lines with wavelength coverage, although gives a larger sample size of $\approx$ 1450 galaxy--absorber pairs), and provides more information than considering all components at each redshift together as a single absorption system (which results in a sample with $\approx$ 90 galaxy--absorber pairs, and treats a single weak line the same as a strong, many-component system). This sample with absorbers `counted' using the maximum number of components is the sample utilized for our analysis in Section \ref{sec:tests}.

Due to blending, it is often difficult to measure the equivalent width (EW) of individual components, so where EW measures are used we measure the total EW of Mg~II (2796 \AA), Fe~II (2382 \AA) or C~IV (1548 \AA) for that absorption system, comprised of many components. These are the EW values listed in Table \ref{table:absorbers}, and shown in our analyses in Section \ref{sec:sample_properties} and Section \ref{sec:metallicity}.

We do not generally attempt to associate individual absorption components or absorption systems with a single galaxy. Given the depth of our spectroscopic data, there are often multiple galaxies with a range of masses that are candidates for associating with absorption features.

\subsection{Sample Depth} \label{sec:completeness}

It is necessary to evaluate the depth of our galaxy and absorber samples, in order to estimate the likelihood of absorption being associated with faint galaxies that are not included in our sample, and of galaxies producing weak absorption that we do not detect. The detection limit of our absorber sample is fairly consistent across the redshift range covered by any ion, leading to a consistent sample discussed further in Section \ref{sec:sample_properties}. Our galaxy sample depends more strongly on redshift due to various emission lines covered by the MUSE and HST spectroscopic data (see Figure \ref{fig:z_dists}), as well as the need for the source to be detected in the F140W imaging. We therefore measure the source density of our sample as a function of magnitude, and compare with deep surveys from the literature.

\begin{figure*}
\includegraphics[trim={0.7cm 0.2cm 0.7cm 0.2cm},clip, width=\textwidth]{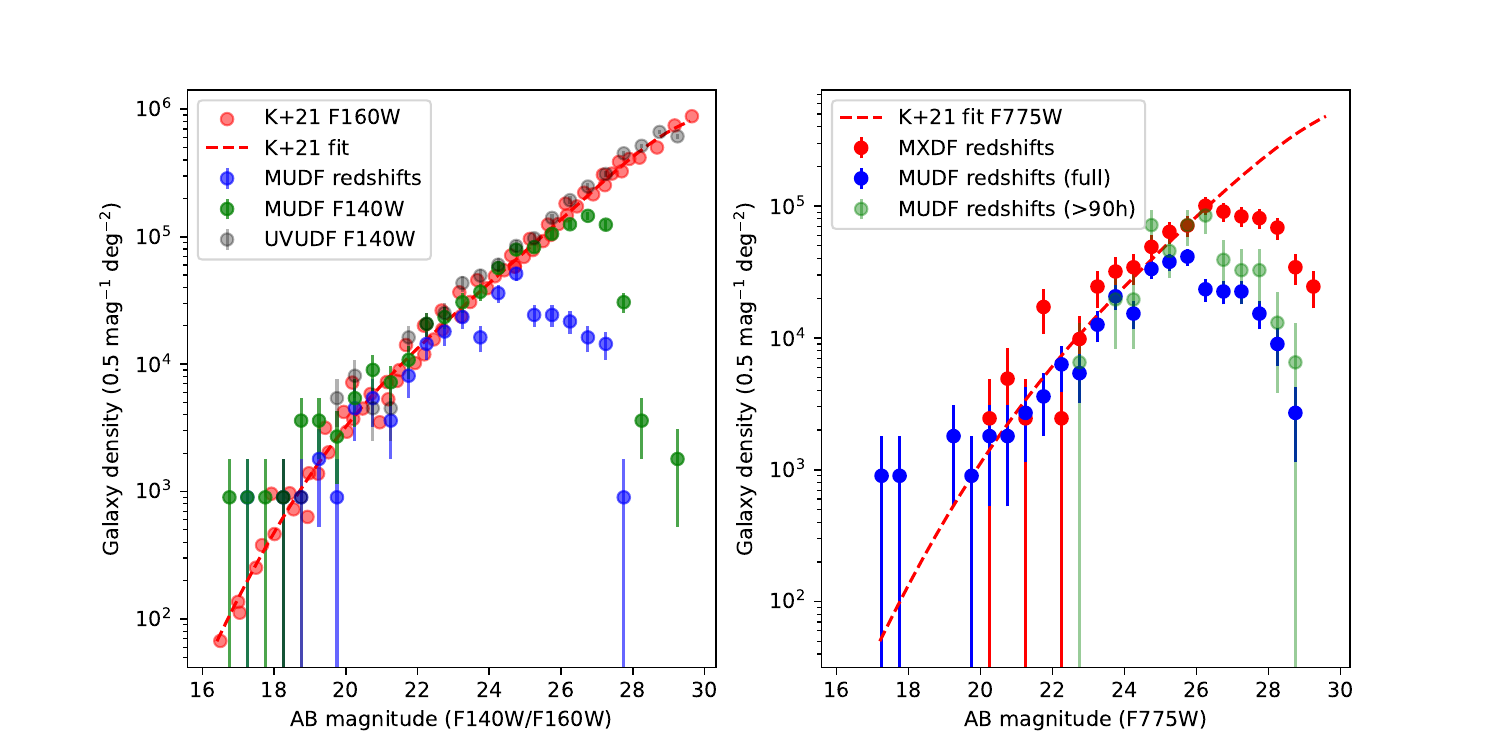}
\caption{Galaxy number counts per square degree as a function of magnitude, comparing the MUDF with other surveys. Poisson errors are also indicated. \textit{Left:} The photometric completeness of the MUDF F140W imaging (from which our spectra are extracted) in comparison with the F160W results shown in Figure 1 of \citet{koushan2021} and the F140W UVUDF imaging \citep{rafelski2015}. The number of sources with confirmed redshifts z $<$ 3.2 is also shown for comparison. \textit{Right:} The number of spectroscopic redshifts z $<$ 3.2 in the MUDF and the MXDF \citep{bacon2023}, as a function of F775W magnitude extracted from the MUSE spectra. We also show the contribution from the central area of the MUDF, where the depth of the MUSE data exceeds 90h.
\label{fig:mag_counts}}
\end{figure*}

This comparison is shown in Figure \ref{fig:mag_counts}. The left panel compares our photometric catalog (green) with results from \citet{koushan2021} and from the UVUDF \citep{rafelski2015}, both of which feature substantially deeper IR imaging than the MUDF (shown in red and grey respectively). The \citet{koushan2021} and UVUDF results are very similar, and our MUDF results are consistent with these deeper surveys for sources brighter than $\approx$27th magnitude. This matches the photometric completeness stated in \citet{revalski2023} for the MUDF. 

Our redshift completeness (shown in blue in Figure \ref{fig:mag_counts}) can be seen to drop substantially below 25th magnitude in F140W. This is due to the difficulty of measuring the redshifts of galaxies without emission lines, especially for faint galaxies without the strong continuum required for detecting absorption features in the spectra. Whilst many of these sources do have photometric redshift estimates, these are not sufficiently accurate for us to associate with the quasar absorption features, so these galaxies cannot be included in our analysis. 

As discussed in Section \ref{sec:redshifts}, our requirement of a detection in F140W excludes most galaxies fainter than m$_{AB} \gtrsim 28$, even if they have emission lines visible in MUSE. As most galaxies are brighter in F140W than at optical wavelengths, these galaxies usually have no continuum detection in any of our HST bands. Continuum detections in the HST imaging are required in order to measure galaxy orientations (required for the analysis in Section \ref{sec:tests}) and masses (required for Section \ref{sec:metallicity}), so any galaxies excluded by this use of F140W to extract spectra could not be used in this work. 

Of galaxies with detections brighter than m$_{AB} = 28$ in F140W, we measure robust redshifts for $\approx44\%$. This level of completeness appears slightly lower than the $\approx$ 50\% estimated for the similarly-deep MUSE eXtremely Deep Field (MXDF, \citealt{bacon2023}), but as they use the F775W band and we use the F140W band to measure completeness, these are not directly comparable.

We show a fairer comparison in the right-hand panel of Figure \ref{fig:mag_counts}. We compare the density of sources for which we obtained redshift measurements with those from the MXDF data. Both the MUDF and MXDF reach $\approx$140h depth in the center of their fields. Magnitudes for both fields are obtained by integrating the MUSE spectra through an F775W filter, and sources with z$>$3.2 are excluded from this comparison as they are not relevant for our study. We also show a fit to the F775W number counts from \citet{koushan2021}, which matches the MUDF and MXDF results for sources brighter than magnitude 25. (This is expected, as our exclusion of z$>$3.2 sources will impact the comparison at m$>$25.) It is clear that the MXDF has higher redshift completeness, visible in the figure and well outside the expected uncertainty due to cosmic variance. In total we find $\approx$220 sources per square arcminute in the MXDF, and only $\approx$100 in the MUDF. 

There is a contribution due to the layout of the exposures in the two surveys. Only $\approx$15\% of the MUDF area has MUSE data exceeding 90h, whereas $>$60\% of the MXDF area exceeds 90h. Figure \ref{fig:mag_counts} also shows the source density for the region in the center of the MUDF where the exposure time exceeds 90h. This matches the MXDF out to magnitude $\approx$26.5, but still only reaches $\approx$140 sources per square arcminute, so does not fully explain the difference between the two surveys.

The remaining difference in the number of detected faint sources is most likely due to the more stringent requirements for us to confirm redshifts. This includes our requirement for a detection in F140W (leading to missed sources at m$\gtrsim$27), and the need for two detected lines to ensure correct identification. Combined, these effects allow a different level of completeness between the MUDF and MXDF despite a very similar maximum depth.

\begin{figure}
\includegraphics[trim={0.4cm 0.1cm 0.8cm 0.8cm},clip, width=\columnwidth]{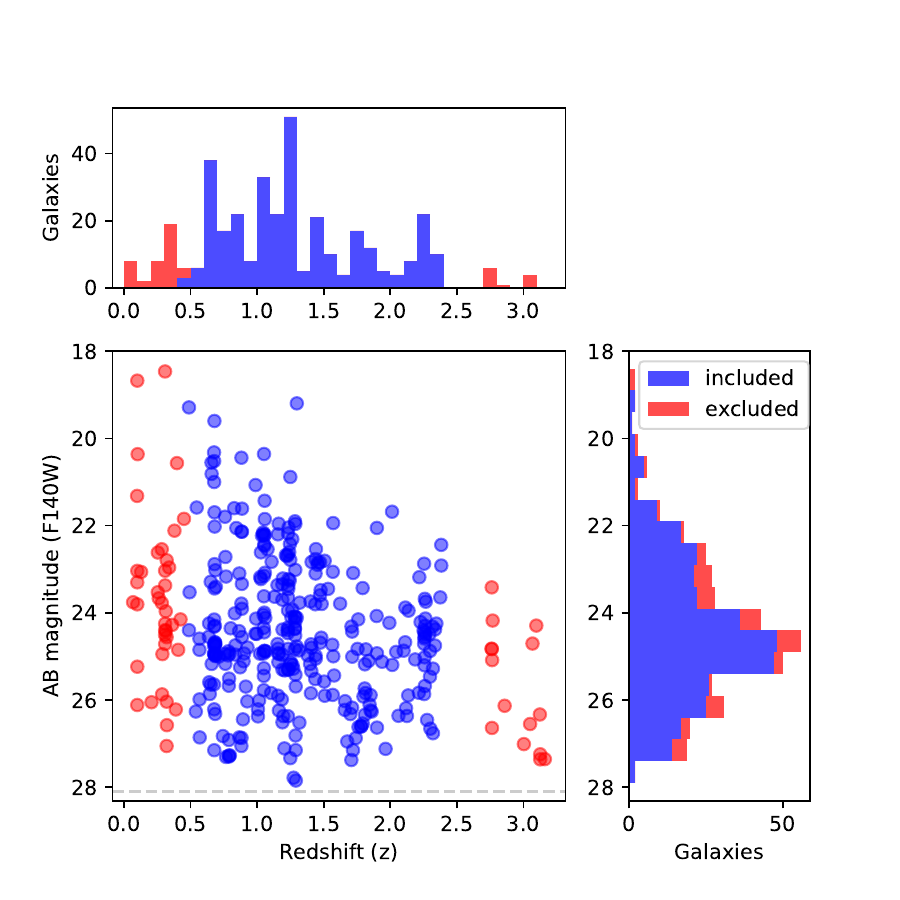}
\caption{Magnitude in the HST F140W imaging against redshift for sources with confirmed redshifts. Sources shown in blue lie in the redshift range 0.45 $<$ z $<$ 2.4, where we have coverage of strong absorption lines in the quasar spectra and strong emission lines in the galaxy spectra, as shown in Figure \ref{fig:z_dists}. Sources outside this redshift range, shown in red, are excluded from most of our analysis.
\label{fig:mag_z}}
\end{figure}

Figure \ref{fig:mag_z} shows the magnitude and redshift of our sources with confirmed redshifts, highlighting the region from 0.45 $<$ z $<$ 2.4 where we have good coverage of emission and absorption lines. This indicates a fairly consistent faint-end limit up to z $\approx$ 2.4, suggesting that our completeness is similar at z$<$1.5 (where redshifts are based primarily on the MUSE data) and at 1.5$<$z$<$2.4 (where they are based on the grism data). We also apply a K-S test to the magnitude distributions at z$<$1.5 and 1.5$<$z$<$2.4 for sources fainter than 24th magnitude in F140W (thereby avoiding effects due to the lack of bright sources at high redshift). This K-S test does not find an inconsistency in the two samples, further supporting a consistent redshift completeness up to z$\approx$2.4. Beyond this, our lack of emission line coverage makes redshift measurements much more difficult. 

For this reason, our analysis in Section \ref{sec:tests} excludes sources at z $>$ 2.4, as we have highly incomplete coverage of the surrounding galaxies. Galaxies below z$\approx$0.45 are also excluded by the lack of strong absorption lines covered by the UVES data at these redshifts. These excluded sources are highlighted in Figure \ref{fig:mag_z}.

In part because of this incompleteness in our galaxy redshift measurements, only 21 of our 30 metal absorption systems feature galaxies at the same redshift, and only 6 of these also allow for galaxy metallicity measurements. This study therefore does not utilize the full potential of the MUDF datasets, and the resulting lack of sample size limits the strength of our conclusions. Observations using the \textit{James Webb Space Telescope} (JWST) will unlock this potential, allowing galaxy metallicities across the full redshift range to be measured.

\subsection{Sample Properties}\label{sec:sample_properties}

Our galaxy--quasar pairs cover a wide range of impact parameters out to $\approx$ 800 kpc, similar to the sample in \citet{beckett2021}; the galaxy--absorber pairs (where absorption is found within 500 \kms{ }of the galaxy) have a median impact paramater of $\approx$380 kpc. The galaxies we measure at 2 $\lesssim$ z $\lesssim$ 3 have masses similar to the z $\lesssim$ 1 galaxies in that work, with measurements down to $10^{8}$ M$_{\odot}$. Moreover, at redshifts $\lesssim$1.5 we can measure the properties of much lower-mass galaxies, approaching $10^{6}$ M$_{\odot}$.

\begin{figure*}
\includegraphics[trim={0.2cm 0.1cm 0.2cm 0.2cm},clip, width=\textwidth]{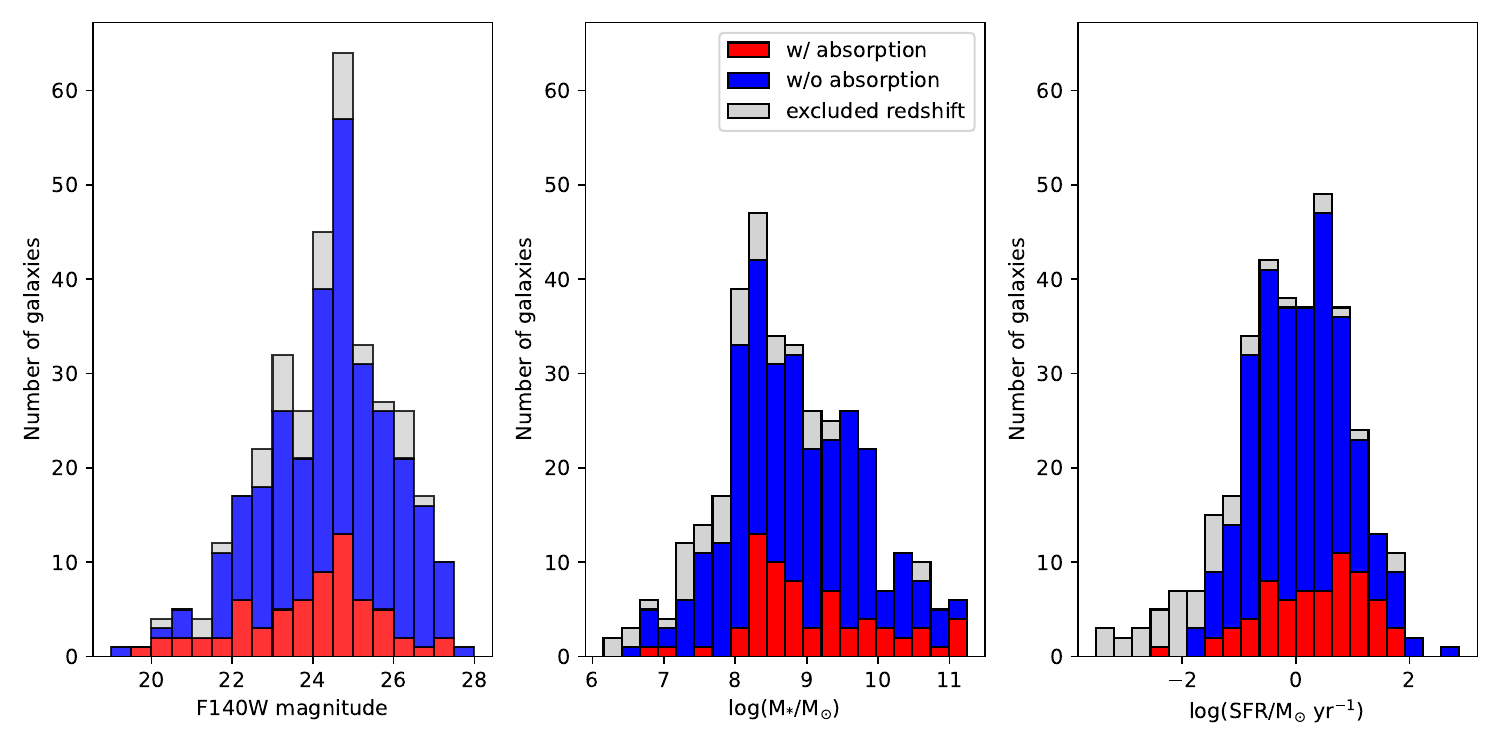}
\caption{Observed magnitudes, stellar masses and star-formation rates of our sample of galaxies. Red bars indicate galaxies for which absorption is found within 500 \kms{ }of the galaxy redshift; blue bars indicate those without detected absorption, but which lie at redshifts 0.45 $<$ z $<$ 2.39 for which the strong Mg~II, Fe~II or C~IV lines are covered by the UVES spectra and multiple emission lines are available for galaxy redshift measurements. In the left panel, we also show the magnitudes of sources with redshifts outside the range for which we could detect absorption (and are hence excluded from our samples), and all sources in the photometric catalog provided in \citet{revalski2023} that are brighter than 28th magnitude.
\label{fig:main_seq}}
\end{figure*}

We illustrate the range of magnitudes, masses and SFRs covered by our sample in Figure \ref{fig:main_seq}. A detailed discussion of the `main sequence' relationship between stellar mass and SFR at low galaxy masses using this data is presented in Fossati et al. (in prep). 

For the purposes of this work, we compare the mass and SFR distributions of galaxies that lie within 500 \kms{ }of detected absorption, and those that do not. A Kolmogorov-Smirnov test \citep[K-S test][]{masseyjr.1951} suggests that the mass distributions of the galaxies with absorption do not differ significantly from those without. The SFR distributions do differ significantly (2.8 $\sigma$ using the K-S test); we find the median SFR for galaxies with absorption is $\approx$0.4 dex higher than for those without. However, we do not dectect a difference between the specific star-formation rates (sSFRs). 

This does not appear to be a direct result of a redshift difference between the samples. Although the distribution with absorbers contains fewer, larger 'peaks', the median redshifts of galaxies with and without absorption are similar. This difference could be an indication that star formation is a factor in producing the absorption around galaxies, but we investigate this further in later sections of this paper.

\begin{figure*}
\includegraphics[trim={0.1cm 0.1cm 0.1cm 0.05cm},clip, width=\textwidth]{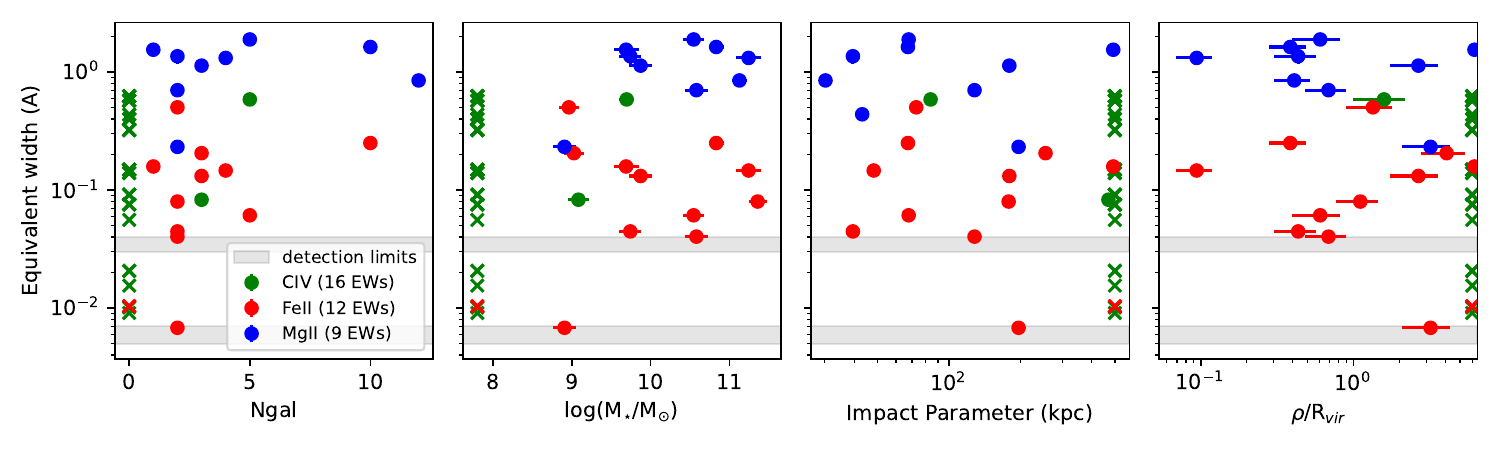}
\caption{The equivalent width (EW) of absorption systems seen along the two lines-of-sight, against various descriptors of the galaxies within 500 kpc and 500 \kms{ } along the line-of-sight. These are: \textit{Left:} the number of detected galaxies; \textit{Center-left:} the total galaxy stellar mass of detected galaxies; \textit{Center-right:} the impact parameter to the nearest galaxy in projected distance;  \textit{Right:} the impact parameter normalized by the galaxy virial radius, for whichever galaxy within 500 \kms{ }is closest by this measure. Points are colored by ion detected in absorption. Absorbers for which no galaxies are detected within this region are shown by crosses, also colored by ion (note that their locations do not represent detection limits on mass or impact parameter). The grey bands indicate the approximate detection limits for the two quasars. We also note that, due to the asymmetry of the survey field, there is an additional source of incompleteness beyond impact parameters of $\approx$ 250 kpc.
\label{fig:abs_ews}}
\end{figure*}

\begin{deluxetable*}{ccccccccccc}
\vspace{1em}
\tabletypesize{\normalsize}
\tablecaption{\normalsize Summary of galaxy properties and locations with respect to the quasar sightlines.}\label{tab:gal_data}
\tablehead{
& & & & & \multicolumn{3}{c}{QSO-NW/faint} & \multicolumn{3}{c}{QSO-SE/bright}  \\
\colhead{ID} & \colhead{z} & \colhead{RA} & \colhead{Dec} & \colhead{\textit{i}} & \colhead{Det} & \colhead{$\rho$} & \colhead{$\alpha$} & \colhead{Det} & \colhead{$\rho$} & \colhead{$\alpha$} \\
\colhead{} & \colhead{} & \colhead{(h:m:s)} & \colhead{(d:m:s)} & \colhead{$^{\circ}$}  & & \colhead{(kpc)} & \colhead{$^{\circ}$} & & \colhead{(kpc)} & \colhead{$^{\circ}$} \\
(1) & (2) & (3) & (4) & (5) & (6) & (7) & (8) & (9) & (10) & (11)
}
\startdata
958 & 0.6763 & 21:42:26.697 & -44:20:14.538 & 48 $\pm$ 20 & N & 480 & 18 $\pm$ 2 & Y & 76 & 46 $\pm$ 2\\
20570 & 0.6767 & 21:42:27.771 & -44:20:05.265 & 40 $\pm$ 3 & N & 508 & 52 $\pm$ 2 & Y & 182 & 66 $\pm$ 2\\
498 & 0.6769 & 21:42:21.932 & -44:20:43.895 & 47 $\pm$ 1 & N & 539 & 75 $\pm$ 1 & Y & 354 & 49 $\pm$ 1 \\
902 & 0.6769 & 21:42:28.033 & -44:20:12.370 & 41 $\pm$ 5 & N & 553 & 29 $\pm$ 1 & Y & 181 & 77 $\pm$ 1 \\
964 & 0.6770 & 21:42:25.394 & -44:20:14.974 & 43 $\pm$ 3 & N & 414 & 90 $\pm$ 1 & Y & 39 & 76 $\pm$ 1 \\
1324 & 0.6773 & 21:42:22.243 & -44:19:51.223 & 54 $\pm$ 5 & N & 156 & 74 $\pm$ 2 & Y & 340 & 22 $\pm$ 2\\
\enddata 
\tablecomments{Galaxy properties are given for those galaxies with spectroscopic redshifts within 500 \kms{ }of detected absorption in either of the two quasar sightlines. Galaxies are sorted by redshift. We show example data from a small number of galaxies here; the remainder are included in machine-readable form. Columns are as follows: (1) Galaxy ID (matching released catalogs from \citealt{revalski2023}); (2, 3, 4) Galaxy redshift and on-sky co-ordinates; (5) galaxy inclination; (6, 9) whether absorption is detected within 500 \kms{ }of the galaxy for the fainter/brighter quasar respectively; (7, 10) impact parameter to the quasar; (8, 11) azimuthal angle between the galaxy major axis and the line joining the galaxy and quasar.}
\vspace{-1em}
\end{deluxetable*}

We can use this MUDF dataset to study the equivalent width of absorption as a function of the number and distance of galaxies at similar redshifts. In Figure \ref{fig:abs_ews}, we show the total equivalent width (EW) of our absorption systems against various measures of galaxy environment within 500 \kms. These include the number of detected galaxies within 500 kpc, the summed stellar masses of these galaxies, and the projected distance to the nearest galaxy, both in physical distance and when normalized by galaxy virial radius. The EWs of absorption systems with no detected galaxies are also shown.

We expect to find an anti-correlation between equivalent width and impact parameter alongside a correlation between equivalent width and number/mass of nearby galaxies, as seen in several previous studies \citep[e.g.][]{nielsen2018, lundgren2021, dutta2021, huang2021a}. 
We test these correlations using a Pearson-r test for each of these three ions. We find that only the correlation between Fe~II absorption and number of nearby galaxies is significant at the 2$\sigma$ level. No correlations are significant if we instead use a Spearman rank test.

Correlations in Mg~II are expected to be weaker, as they are likely driven by the physical column density. Many of these absorption systems have velocity widths $\sim$ 50-100 \kms. At this width, the absorption lines begin to saturate at $\sim$ 1-2 \AA, affecting our stronger absorbers such that EW reflects the gas kinematics rather than the column density. 

We note that most studies do not find absorption with these large EWs at impact parameters $>$100kpc. This may be a result of completeness effects. As mentioned in Section \ref{sec:galaxies}, we cannot accurately determine redshifts for most quiescent galaxies, nor do we include the less-confident redshifts from those exhibiting only a single emission line. Our absorbers could therefore be assigned a large impact parameter because we are unable to measure a redshift for a closer galaxy. This is unlikely near the center of our field, as the depth of our MUSE data suggests that our sample is likely more complete than most previous work. However, we could be affected by incompleteness due to the edge of the field-of-view and the shallower spectroscopic data near the field edge. 

\citet{dutta2020} also find a small number of absorbers with large EW ($\approx$ 1 \AA) and large impact parameter ($>$200 kpc) in galaxy groups. With our small sample size, only one or two absorbers with elevated EW due to the presence of a group could mask any anti-correlation. Galaxy misassociation (i.e. if the absorbing gas is due primarily to one galaxy, it may not be the one selected), due to either incompleteness or the existence of a galaxy group, would likely weaken any of the expected correlations, so we cannot detect them in our sample of absorbers. Our C~IV absorbers lie at higher redshifts where fewer emission lines are visible, so there are fewer detected galaxies and no correlation can be found.

\begin{figure*}
\includegraphics[trim={0.0cm 0 0 0},clip, width=\textwidth]{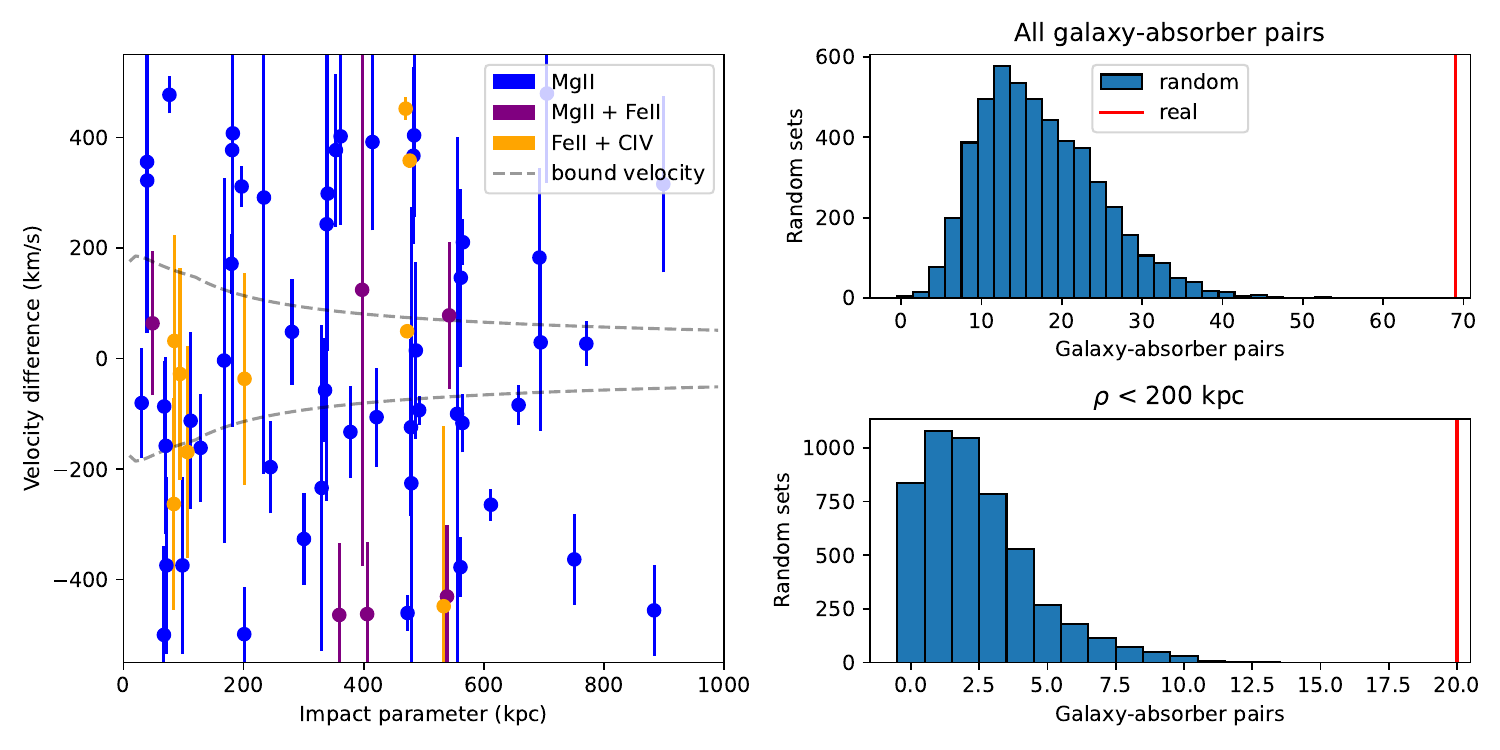}
\caption{The number of galaxies in our sample within 500 \kms{ } of detected absorption, as a function of velocity offset and compared with randomly-placed absorbers. \textit{Left:} Galaxy--absorber velocity offset against impact parameter for pairs in our sample. Colors indicate the ions detected. Uncertainties shown reflect the spread of absorption components and the galaxy redshift uncertainties. We also show the velocity at which gas becomes unbound from a typical galaxy of stellar mass 10$^{9.5}$ M$_{\odot}$ at z $\approx$ 1.0 (without transverse velocities and line-of-sight distances this is only indicative). \textit{Right:} The number of galaxies found within 500 \kms{ } of absorption when absorber redshifts are randomized as described in Section \ref{sec:sample_properties}. The histogram in the upper panel shows the distribution for 5000 sets of absorbers with randomized redshifts, whilst the red line shows the number of galaxies found with absorption in the real MUDF data. The lower panel is similar, but only considers galaxy--absorber pairs with impact parameters smaller than 200 kpc.
\label{fig:dv_imp}}
\end{figure*}

The velocity offset between galaxies and absorbers in these galaxy--absorber pairs are also of interest, as it can illustrate whether absorbers are likely bound to the galaxy halo. We show this distribution in Figure \ref{fig:dv_imp}, alongside an indication of which absorbers are likely to be unbound from the galaxy. Due to the lack of transverse velocity measurements and line-of-sight radii, both distances and velocities are likely to be larger than shown, and hence less likely to be bound. 


The relative lack of absorbers at small impact parameters and small velocity offsets from galaxies raises the question of whether these absorbers are physically associated with the nearby galaxies, or just random placement of absorbers near galaxies. We confirm this by measuring the expected number of galaxies around our absorbers if the absorbers were randomly placed. We perturb the redshifts of our absorbers randomly by up to dz = 0.3 in either direction, whilst ensuring that they remain within the 0.45 $<$ z $<$ 2.39 region where we have wavelength coverage of Mg~II/Fe~II absorption and galaxy redshifts from multiple emission lines.
We then count the number of galaxies that would be associated with absorbers at these randomized redshifts. The right-hand panels of Figure \ref{fig:dv_imp} show the distribution of the number of galaxy--absorber pairs in each of 5000 sets of randomized absorbers. The real MUDF features more galaxy--absorber pairs than any of the randomized sets, indicating that a large fraction of the pairs in our sample are indeed physically associated. Based on these results, we expect $\approx$ 75\% of the galaxy--absorber pairs to be real associations, as these are not found in the randomized pairs. Requiring impact parameters $<$200 kpc increases the likely fraction of real associations to $\approx$ 90\%, but also greatly reduces the sample size.

Given the fairly conservative velocity cut and the large impact parameters in our sample, it is unsurprising that a many of our galaxy--absorber pairs are likely unbound. This does not necessarily mean that the absorber and galaxy are physically unrelated, as outflows from low-mass galaxies must reach large distances in order to enrich the IGM \citep[e.g.][]{booth2012}. Some outflows are expected to reach several hundred \kms{ }in the CGM across our redshift range \citep[as suggested by observations and simulations, e.g.][]{nelson2019, mitchell2020, schroetter2019}, although may slow at larger impact parameters.

Additionally, our absorption systems are made up of many components at slightly different velocities, often covering a range 100 \kms or larger. In these cases it is possible that only some components are bound and/or associated with a particular galaxy. In order to allow for these additional velocities and to ensure we do not remove outflows from our sample, we do not exclude galaxy--absorber pairs based on the likelihood that they are unbound (other than our initial cut at 500 \kms).

\section{Tests of position angle bimodality} \label{sec:tests}

\subsection{Methods}

In this section we consider the azimuthal angle distribution of galaxy--absorber pairs in the MUDF data, to determine whether this distribution also exhibits the bimodality seen in H I in the Q0107 system \citep{beckett2021}, and in other Mg~II studies \citep[e.g.][]{bouche2012, kacprzak2012, schroetter2019}. This uses the Hartigan dip test \citep{hartigan1985}, which measures the maximum difference between the empirical distribution and the unimodal distribution that minimizes said difference.
In this case, the deficit of absorption at intermediate angles is used to determine whether there is a significant excess of absorption along the projected major and minor axes of galaxies.

This is preferred over other common tests of multimodality \citep[e.g bandwith tests,][]{silverman1981}, as these may be distorted by the large difference in azimuthal angle between the expected modes, or by the bounds of the distribution at 0 and 90$^{\circ}$. Other difference tests (e.g. Kolmogorov-Smirnov,  \citealt{masseyjr.1951}) could be affected by a single mode if calculated against a uniform distribution, so may not directly test for the bimodality expected due to gas recycling.

As discussed in Section \ref{sec:sample_construction}, our sample of galaxy--absorber pairs is selected by combining galaxies for which spectroscopic redshifts have been measured (as described in \citealt{revalski2023}) with any detected absorbers within 500 \kms{ } in either of the observed quasar lines-of-sight. This inclusion of all galaxies around each absorber means that we do not need to determine the `true' associated galaxy, but means that each absorber is paired with an average of 4 galaxies. This adds noise resulting from 'false' associations to the azimuthal angle distribution. This noise should be uniform but may `wash out' some of the bimodal `signal'.

In order to ensure a consistent test across redshifts, we only consider the Mg~II and Fe~II absorption. As these have similar ionization potentials, they are expected to probe similar gas phases with temperatures $\approx 10^{4}$K, whereas C~IV generally probes warmer material nearer to $10^{5}$K \citep[e.g.][]{gnat2007}. The number of absorbers at that location is the number of components found in that sightline for the ion with the most components. For this test we only consider galaxies with inclinations greater than 20$^{\circ}$, as accurate position angle measurements could not be obtained for galaxies with very low inclinations (see Appendix \ref{sec:pa_measurements}).

We compare with the results from \citet{beckett2021} covering the Q0107 field. As with the MUDF, the quasars were selected due to the presence of multiple lines-of-sight, so the foreground absorption is not biased. The galaxy surveys also both cover a wide field, allowing us to access impact parameters larger than a single MUSE pointing. Unlike \citet{beckett2021}, our absorber sample does not feature H I absorbers, rather absorbers from metal transitions generally probing cool gas.

Azimuthal angles are measured between the projected major axis and the line from the galaxy center to each quasar. These are collapsed to the interval from 0 to 90$^{\circ}$, with 0$^{\circ}$ showing absorption along the major axis, and absorbers at 90$^{\circ}$ along the minor axis (i.e. the direction of the azimuthal angle does not matter). We note that the distribution of azimuthal angles for all galaxies is close to uniform, so any structure shown in the distribution of galaxy--absorber pairs is due to the presence of detected absorption, not the underlying distribution of galaxy orientations.

In order to account for the effect of changing sample size on the results of the Hartigan dip test, as well as provide an estimate of the uncertainties on the p-values returned by the test, we use a resampling procedure. For each sample of galaxy--absorber pairs, we draw an ensemble of bootstrapped samples from the PA distribution of sizes between 20 and 800 detections. The median and percentiles can then be used to track how the test results and their robustness changes with sample size. We find that ensembles of 50 bootstrapped samples produce medians and 1$\sigma$ uncertainties that are consistent enough for this test. The same procedure is used to estimate the uncertainties on our results by producing bootstrapped samples with the same number of detections as the observed sample. We can therefore discuss any detection of a significant bimodality at the 2$\sigma$ level (p $\lesssim$ 0.05), as well as the robustness of any such bimodalities found.

\subsection{MUDF Results}

\begin{figure}
\includegraphics[trim={0.1cm 0 0 0},clip, width=\columnwidth]{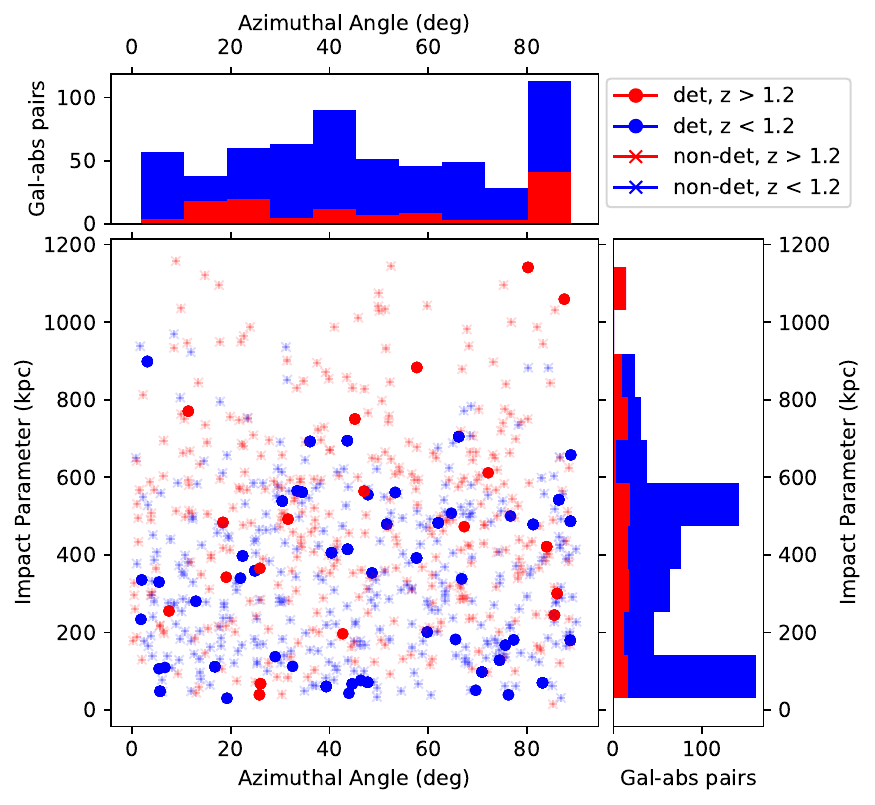}
\caption{Azimuthal angle against impact parameter for galaxy--sightline pairs in the MUDF sample. Where absorption is detected within 500 \kms, we show solid points colored by redshift, with non-detections shown as faint crosses. Only the detections are projected into histograms, also coloured by redshift. Note that multiple absorption components are often found for the same galaxy and sightline, leading to one point on the scatter plot corresponding to multiple detections in the histograms.
\label{fig:pa_inc_mudf_all}}
\end{figure}

\begin{figure}
\plotone{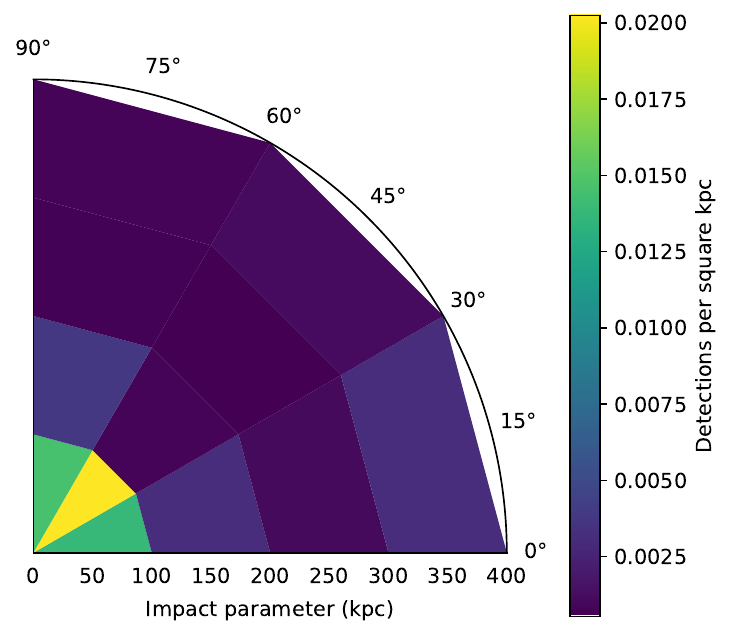}
\caption{Azimuthal angle against impact parameter for galaxy--absorber pairs in the MUDF sample, split into 30$^{\circ}$ by 100 kpc bins. These are normalized by area to account for the physically larger bins at larger impact parameters, but not normalized to account for the partial coverage at $\rho \gtrsim$ 300 kpc due to the non-circularity of the survey field.
\label{fig:pa_polar_all}}
\end{figure}

We show the impact parameters and azimuthal angles of galaxy--absorber pairs with velocity separation $<$ 500 \kms{ }in Figure \ref{fig:pa_inc_mudf_all} (bold points), as well as galaxy--sightline pairs for which we do not detect absorption (faint crosses). This shows that only a small fraction of galaxy--sightline pairs feature absorption ($\approx$15\%). The detections are projected into histograms to show the azimuthal angle and impact parameter distributions.
A bimodal distribution in azimuthal angle is not immediately apparent, and indeed a dip test produces a result of p $\approx$ 0.2, a non-detection at the 2$\sigma$ level. This differs from the Q0107 sample of galaxy--absorber pairs described in \citealt{beckett2021} (which covers a similar range of impact parameters out to $>$ 500 kpc), as well as other studies with smaller samples \citep[e.g.][]{bouche2012, schroetter2019}, for which the bimodality is clear and significant within the innermost 100 kpc. 

We note that the histogram of impact parameters shows that both the innermost bin and another at $\approx$ 500 kpc show a far larger number of detected galaxy--absorber pairs than other bins. Both bins include galaxy--absorber pairs at a range of redshifts between z=0.5 and z=1.5 (they are not dominated by a single galaxy group or absorption system). It therefore appears that the excess in the innermost bin is due to more absorption found near to galaxies. The 500 kpc bin represents a larger area on the sky than those at smaller radii, and there is a contribution from the group at z=0.88, where multiple absorption components in both quasars separated by 500kpc are paired with multiple nearby galaxies. The shape of the field also reduces the number of galaxies at impact parameters $\gtrsim$ 300 kpc.

The impact parameters and azimuthal angles for galaxy--absorber pairs within 400 kpc are then binned in Figure \ref{fig:pa_polar_all}, and normalized by bin area to account for physically larger bins at larger impact parameters. This normalization highlights that the number of detections is much higher in the innermost 100 kpc, but we do not see excess absorption along the major and minor axes. However, we do see a small excess of absorption along the major and minor axes at impact parameters of 100-200 kpc.

\begin{figure*}
\includegraphics[trim={0.1cm 0 0 0},clip, width=\textwidth]{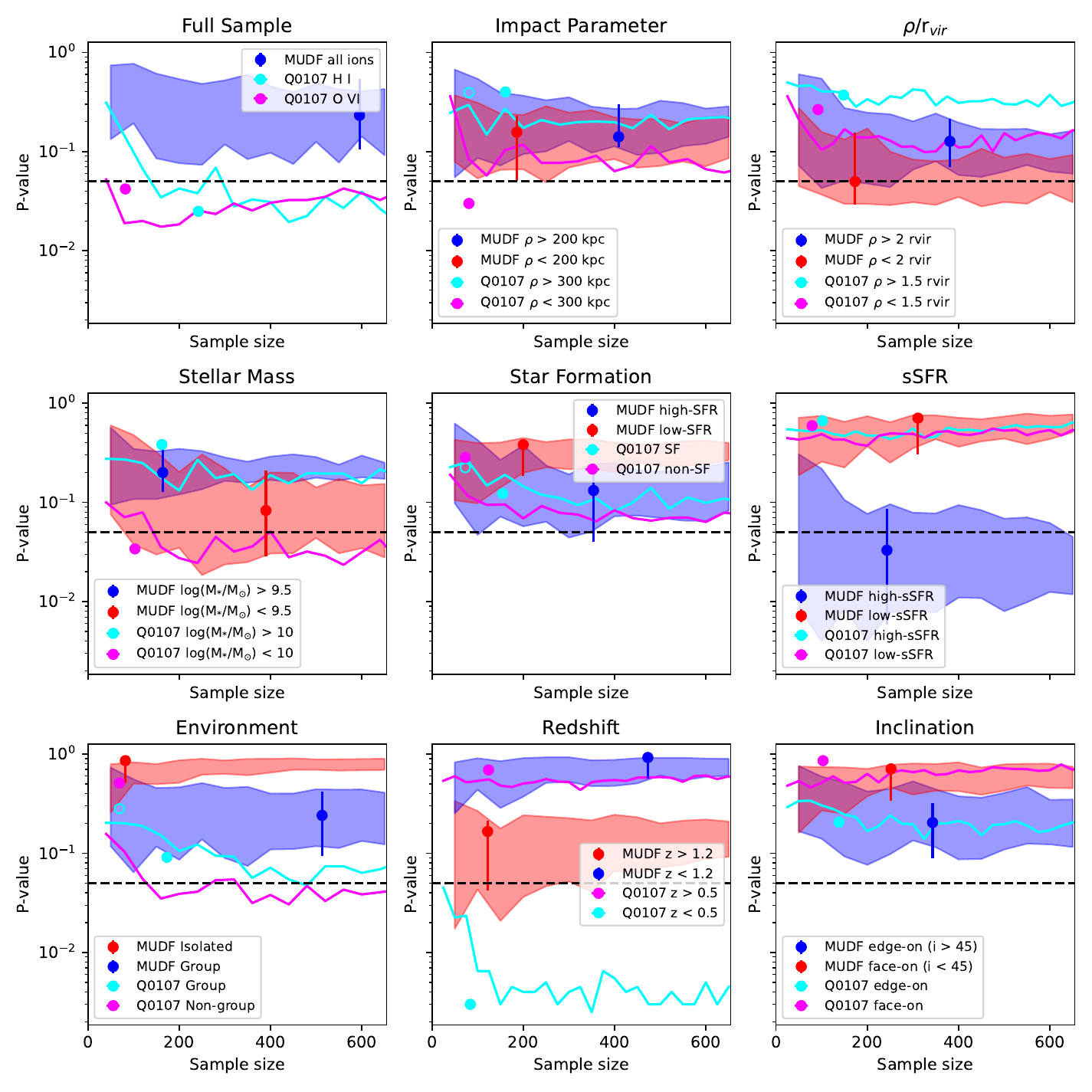}
\caption{The p-values returned from a Hartigan dip test applied to both the MUDF and the Q0107 data \citep{beckett2021}, as a function of the number of galaxy--absorber pairs detected in the sample. The MUDF data are shown by the red and blue points, with uncertainties estimated using our resampling procedure. The regions between the 16th and 84th percentiles calculated using this resampling are shaded. Cyan and magenta points shown are the results given in Section 4 of \citep{beckett2021}, with solid points showing the actual data, and unfilled points showing resampling to match sample sizes. These are also resampled across a range of sample sizes, with the median results shown by the solid cyan and magenta lines. Points below the dashed black line indicate detection of a bimodal distribution at $>$ 2$\sigma$ significance. The details of how each sample is divided are given in the text. 
\label{fig:sample_size_mudf}}
\end{figure*}

In Figure \ref{fig:sample_size_mudf} we separate the sample as a function of galaxy and absorber properties, show how the dip test results are affected by sample size, and compare our results with those from the low-redshift study of the Q0107 field \citep{beckett2021}. The results from the full sample of galaxy--absorber pairs are given in the top-left panel. The results shown in this Figure are also provided in Table \ref{table:PA_results}.

For the full samples, the Q0107 H~I results are mostly below the p=0.05 line, indicating stronger than 2$\sigma$ significance. The MUDF metal-line results, however, remain above this threshold, indicating that we do not confidently detect a bimodality in the azimuthal angle distribution.

The sample sizes for the full MUDF sample and the Mg~II sample are both substantially larger than the full Q0107 sample, and reproduce the results described above when bootstrapped to the size of the Q0107 sample. Therefore, the lack of bimodality does not appear to be due to any direct effect of the sample size. As discussed in Section \ref{sec:data} and Appendix \ref{sec:pa_measurements}, uncertainties or systematic errors in our position angle measurements are also unlikely to have a significant effect.

Given the multitude of differences between the galaxy and absorber samples in this work and those used in B21, there are numerous other possible reasons for the difference between the MUDF and Q0107 results. Possible explanations include evolution in redshift, different metal ions visible in absorption, different galaxy detection limits, and different survey geometry (and hence different impact parameter distribution). In the following sections we investigate such factors and attempt to determine whether a cause of the different results between this work and B21 can be found. As with the Q0107 sample, we split the MUDF sample into a series of complementary subsamples, with the results shown in the remaining panels of Figure \ref{fig:sample_size_mudf}.

\begin{table}
\begin{center}
\caption{Results from applying the Hartigan dip-test to samples from the MUDF. \label{table:PA_results}}
\begin{tabular}{ c c c}
 \hline Sample & Gal--Abs Pairs & p-value  \\\hline
 Full & 595 & 0.232   \\ \hline
 $\rho$ $<$ 200 kpc & 186 & 0.157 \\
 $\rho$ $>$ 200 kpc & 409 & 0.141 \\ \hline
 \textbf{$\rho$ $<$ 2 r$_{vir}$} & \textbf{173} & \textbf{0.048} \\
 $\rho$ $>$ 2 r$_{vir}$ & 381 & 0.127 \\ \hline
 M$_{*}$ $<$ 10$^{9.5}$ M$_{\odot}$ & 390 & 0.081 \\
 M$_{*}$ $>$ 10$^{9.5}$ M$_{\odot}$ & 164 & 0.200 \\ \hline
 Low-SFR  & 200 & 0.382 \\
 High-SFR & 354 & 0.132 \\ \hline
 Low-sSFR  & 311 & 0.707 \\
 \textbf{High-sSFR} & \textbf{243} & \textbf{0.033} \\ \hline
 Isolated (350 kpc/350 \kms) & 82 & 0.858 \\
 Group & 513 & 0.241 \\ \hline
 z $<$ 1.8  & 473 & 0.923 \\
 z $>$ 1.8 & 122 & 0.166 \\ \hline
  20 $<$ i $<$ 45 (face-on)  & 252 & 0.708 \\
 i $>$ 45 (edge-on) & 343 & 0.203 \\ \hline
 
 
\end{tabular}
\tablecomments{Results from applying a Hartigan dip test to various samples drawn from the MUDF catalogues, as shown in Figure \ref{fig:sample_size_mudf}. Columns list a descriptor of the sample, the number of detected galaxy--absorber pairs in that sample, and the p-value resulting from the dip test. Samples with an azimuthal angle bimodality that is significant at the 2$\sigma$ level are bolded.}
\end{center}
\end{table}

\subsection{Investigating MUDF sub-samples}\label{sec:bimodality_subsamples}

We see no bimodality in the distribution when split by impact parameter (upper-middle panel of Figure \ref{fig:sample_size_mudf}). Although we show a split at 200 kpc, no threshold value produces a significant result in either sample. However, when we normalize the impact parameters by the galaxy virial radius, we do see a bimodality at the 2$\sigma$ level in galaxy--absorber pairs closer than 2r$_{vir}$ (upper-right panel). This highlights the importance of accounting for different virial radii when comparing the CGM of different galaxies.

We also split the sample by stellar mass (middle-left panel), star-formation rate (central panel) and specific star-formation rate (middle-right). Only the high-sSFR sample produces a 2$\sigma$ bimodality, although the low-mass and high-SFR samples produce a more plausible bimodality than their counterparts. This is consistent with the bimodality being driven by stellar feedback, producing outflows which can extent to larger distances in shallower gravitational potentials.

We note that for mass, SFR and sSFR we test the sample split using a fixed threshold, and also fitting a 4th-order polynomial to mass/SFR/sSFR as a function of redshift, using this to split the sample whilst accounting for redshift evolution in galaxy masses and SFRs. Stronger results are found when using a fixed stellar-mass threshold (10$^{9.5}$ M$_{\odot}$ is shown), but when using the redshift-dependent fit to select high-SFR and high-sSFR galaxies.

The splits shown in the lower panels of Figure \ref{fig:sample_size_mudf} do not reveal any bimodality, and hence no evidence for the proposed disk and outflow structures. We do not find any split by galaxy environment that produces a bimodal distribution, although we note that our deep spectroscopy means that we have very few isolated galaxies in our sample. In the lower-left panel we split the sample by defining isolated galaxies as those which are the only galaxy within 500 \kms{ }of the absorber redshift and 500 kpc of the sightline, but the results are similar for any linking length used to define a group, and whether pairs are included as `group' galaxies or `isolated' galaxies, or excluded from the sample entirely. We also see no significant bimodality in the azimuthal angle distribution when considering galaxy redshift (lower-middle panel) or inclination (lower-right).

We also check whether these results change when we use a stronger cut in inclination to remove galaxies that are close to face-on (i $<$ 40$^{\circ}$). We find that the reduction in sample size slightly reduces the significance of the bimodalities found in our sub-samples, with the bimodality in the low-impact-parameter sample no longer reaching a 2$\sigma$ significance. 

\begin{figure*}
\includegraphics[trim={0.1cm 0 0 0},clip, width=\textwidth]{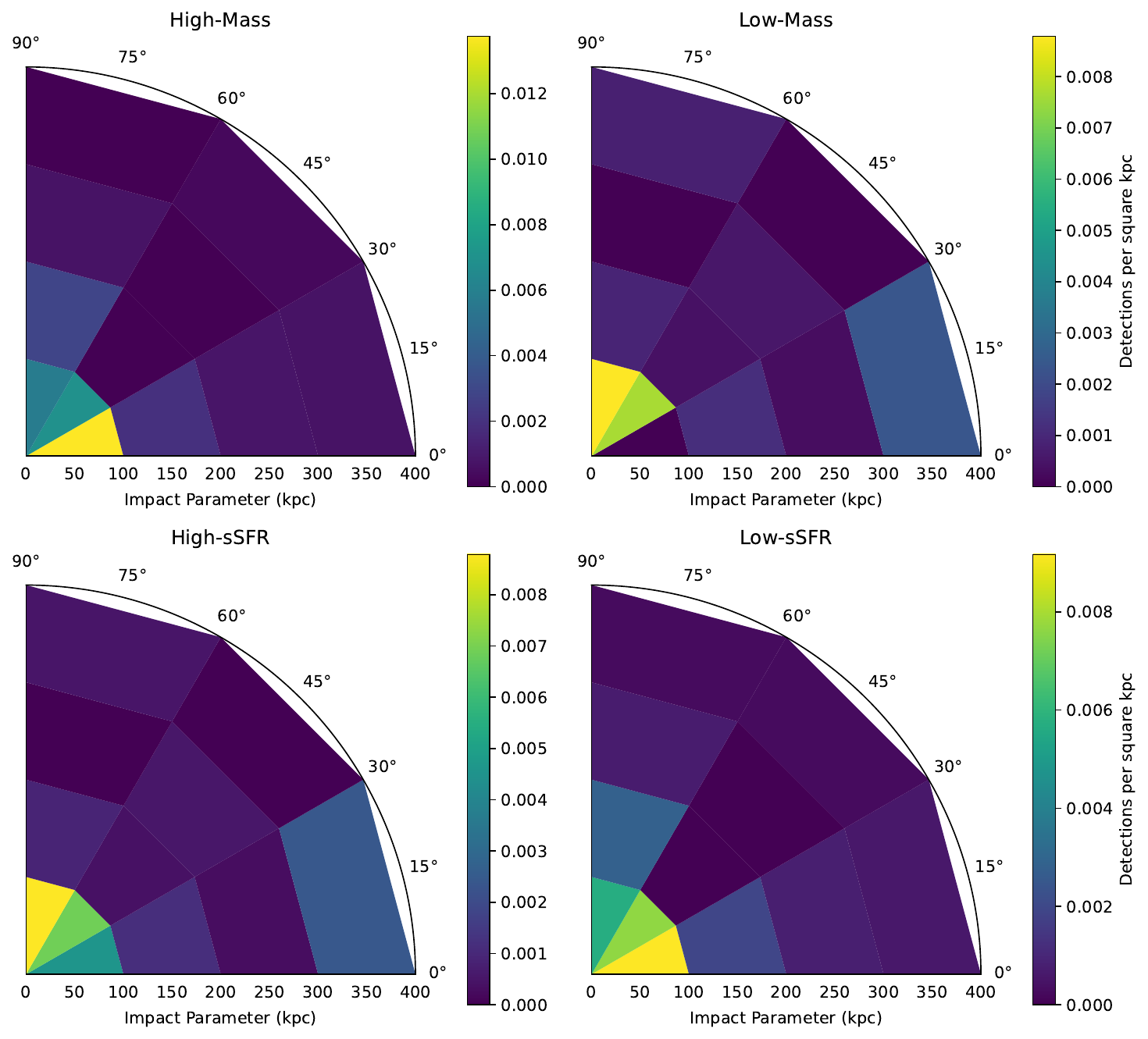}
\caption{Azimuthal angle and impact parameter of galaxy--absorber pairs for galaxies for four of the subsamples shown in Figure \ref{fig:sample_size_mudf}, namely high-mass, low-mass, high-sSFR and low-sSFR. Only the high-sSFR sample exhibits an overall bimodality. These are binned and normalized by projected area as in Figure \ref{fig:pa_polar_all}. Note that the colors in each panel are scaled separately to highlight the innermost bins, so cannot be used to compare detection numbers between the panels.
\label{fig:polar_highsfr}}
\end{figure*}

Both of the subsamples showing a bimodality in their azimuthal angles are only a little over 2$\sigma$ significance. These are not robust under our resampling procedure, and adjusting the threshold values used to split the samples can remove the bimodality. We illustrate the results from the mass and sSFR splits in Figure \ref{fig:polar_highsfr}. Only the high-sSFR sample reaches 2$\sigma$ significance overall, but all four subsamples show a small excess of absorption near the major and minor axes in the 100-200 kpc bin. 

This also shows that a contribution to the bimodality comes from the major axis bin at 300-400 kpc. This bin contains $\approx$40 galaxy--absorber pairs, compared with 5-10 in most surrounding bins in Figure \ref{fig:polar_highsfr}. However, these originate from a range of redshifts and the galaxies are not heavily concentrated in one part of the field, so there is no intrinsic alignment between the galaxies contributing to the galaxy--absorber pairs in this bin and no clear reason to dismiss this contribution to the overall azimuthal angle distributions of these samples.

\subsection{Comparison with Q0107 Results}\label{sec:q0107_comp}

\citet{beckett2021} applied similar splits to their low-redshift H~I galaxy--absorber pairs, allowing for a comparison between low-z H~I and metal absorption extending to higher redshifts. The impact parameters of their galaxy--\textit{quasar} pairs are comparable to our sample, although their H~I sensitivity allows them to detect more absorption at large impact parameters. 

Their results when split by impact parameter also generally fail to produce a significant bimodality, although some threshold values do allow a $\approx 2 \sigma$ result. As with the MUDF, the Q0107 results do not show a bimodal distribution when split by inclination or star-formation rate. They also obtain a similar result from splitting their sample by stellar mass as we do at higher redshifts, although the bimodality in their low-mass sample is slightly stronger than $2 \sigma$ where ours is slightly weaker.

However, there are several larger differences between their results and ours. They do find a difference in absorption between group and non-group galaxies, with non-group galaxies more likely to show a bimodal azimuthal angle distribution. Our result from the high-sSFR sample is not reproduced in their work. Finally, the bimodality they detect in their full sample appears to be mainly driven by the lower-redshift galaxy--absorber pairs in their sample, whereas we do not find a significant bimodality when splitting by redshift. We discuss how we can interpret these results in Section \ref{sec:discussion}.

\subsection{Effects of Spectral Resolution}\label{sec:resolution}

The spectra of the quasars in the background of the MUDF, taken using UVES, have a much higher resolution ($\approx$ 7.5 \kms) than the COS ($\approx$ 20 \kms) and FOS ($\approx$ 100 \kms) spectra of the Q0107 triplet used in \citet{beckett2021}, thus allowing us to resolve a larger number of components at each redshift with absorption. 

In order to ensure that this does not substantially affect our results, we repeat the fitting procedure described in Section \ref{sec:absorption} on a smoothed version of the quasar spectra. This is smoothed using a Gaussian kernel in order to match the width of the COS line-spread function, with random noise added to match the signal-to-noise ratio (SNR) per resolution element of the spectra of the Q0107 A and B quasars (which have very similar SNR). 

We re-fit only the Mg~II features. In most cases this is the ion with the largest number of components in any absorption system, so the effects of this change on the results are accurately measured. 

We find that the large majority ($\approx 90\%$) of components are still recovered using the lower-resolution spectra, losing 7 of 70 Mg~II absorbers through blending. These absorbers contribute a total of $\approx$55 galaxy--absorber pairs, approximately 10 \% of the sample.

This small reduction in sample size therefore has only a small effect on the results, with the high-sSFR and low-impact-parameter samples again the only samples of those shown in Figure \ref{fig:sample_size_mudf} that produce a bimodality at the 2$\sigma$ level, and the low-stellar-mass sample approaching that threshold.

This test suggests that the differing resolution (and signal-to-noise) between the quasar spectra used for the high-z MUDF measurements and the low-z Q0107 data is not the main reason for the different results obtained from their azimuthal angle distributions.

\subsection{Summary of Azimuthal Angle Results}\label{sec:bimodality_summary}

To summarize, the full MUDF sample does not show a significant bimodality, but we do see a result at close to the 2$\sigma$ level in galaxy--absorber pairs with high galaxy sSFR and those with small impact parameters relative to the galaxy virial radius. The low-stellar-mass and high-SFR samples also approach the 2$\sigma$ threshold. We note that these are not generally robust under our bootstrapping procedure or under changes to the thresholds used to divide the samples. If we consider only galaxy--absorber pairs with velocity differences $<$ 300 \kms, the sample size is smaller but the overall results remain similar, with the same subsamples producing a significant, although usually less robust, result.  

For many of these subsamples, we also try removing the galaxy group (at z$\approx$ 1.05), as we would not expect to see a clear bimodality in a large group and the large number of galaxies means that this contributes a larger fraction of the galaxy--absorber pairs in our sample than any other single system ($\approx$ 150 pairs). Removing this group does not allow any additional samples to exhibit a bimodality.

We can therefore report a tentative bimodality around strongly-star-forming galaxies and at impact parameters smaller than 2 r$_{vir}$, but given the small number of galaxy--absorber pairs at low impact parameters we are unable to obtain significant results from any further splitting of the sample.

\section{Relations between absorber properties and galaxy metallicity}\label{sec:metallicity}

We also discuss the association between galaxy metallicity and properties of nearby gas seen in absorption. The galaxy metallicities are discussed in detail in \citet{revalski2024}, here we utilize their results for comparison with the metal absorption lines.

In order to compare across the wide range of redshifts, stellar masses and star-formation rates covered by the MUDF data, we consider the difference between the galaxy metallicity and that expected for a galaxy with the same stellar mass and SFR lying on the fundamental metallicity relation (FMR), as parameterized by \citet{curti2020}. We adopt the changing FMR low-mass slope (denoted $\gamma$) described in \citet{revalski2024} to account for the low SFRs of many of our galaxies.

We note that, as discussed in Section \ref{sec:sample_construction}, galaxy metallicities are only available for a small range of redshifts that only partially overlaps with the redshifts at which absorption is visible. Only two absorption systems are found within 400 kpc of at least one galaxy with measured metallicity, both of these representing galaxy groups (z=0.679 and z=2.253). All the results discussed here are therefore based on a very small sample size.

\begin{figure}
\includegraphics[trim={0.1cm 0 0 0},clip, width=\columnwidth]{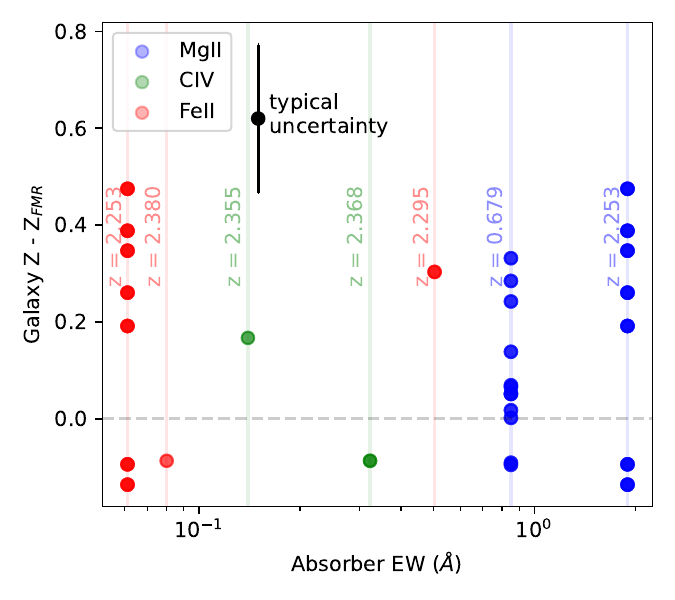}
\caption{Absorber equivalent width (EW) against galaxy metallicity offset from the FMR for galaxy--absorber pairs within 500 \kms. Multiple points along the same vertical lines indicate multiple galaxies with different metallicities close to a single absorber (the redshift of the absorber is labelled). The black point in the top-left indicates typical uncertainties on the metallicity measurement; the uncertainty in EW is generally too small to be visible.
\label{fig:ew_metallicity}}
\end{figure}

We illustrate the sample in Figure \ref{fig:ew_metallicity}, which shows the absorber EW and galaxy metallicity offset from the FMR ($\Delta Z$) of galaxy--absorber pairs separated by $<$ 500 \kms. This illustrates that such pairs only exist for a small number of absorption systems\footnote{Other ions are found in absorption at these redshifts but are not shown for clarity.}. No relationship is found between absorber equivalent width (or any other measured property) and galaxy metallicity. 

\begin{figure}
\plotone{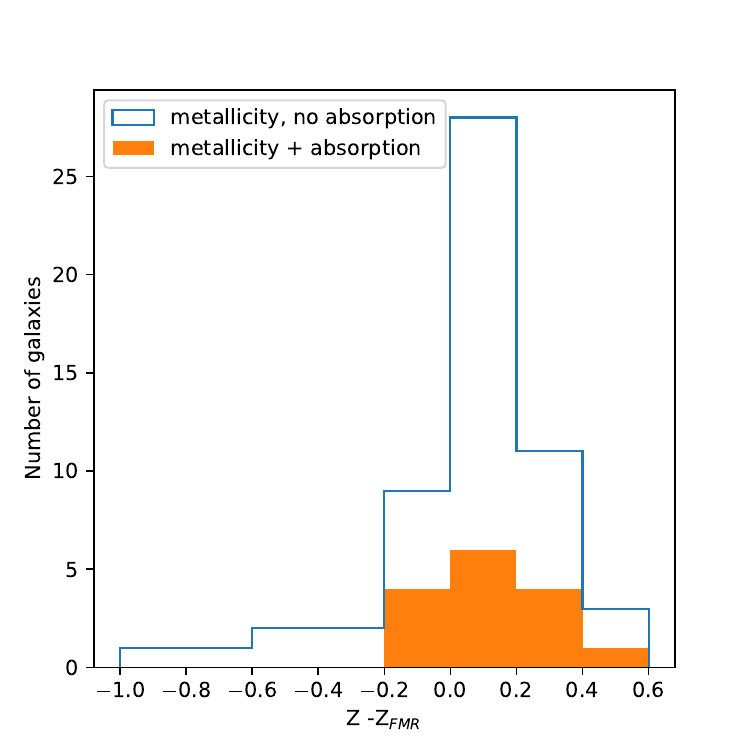}
\caption{The distribution of galaxy metallicities with respect to the FMR at the mass/SFR of the galaxy. The blue histogram indicates all galaxies for which absorption could be detected (i.e. the redshift allows for Mg~II, Fe~II or C~IV to be seen in the UVES spectra), whilst the orange indicates galaxies for which absorption is detected.
\label{fig:FMR_offset}}
\end{figure}

Figure \ref{fig:ew_metallicity} appears to show the majority of galaxies with absorption have higher metallicities than expected using the FMR. In order to test whether this is representative of the full sample of galaxies with metallicity measurements, we compare the distribution of these galaxies with metallicity measurements and nearby absorption and those without absorption. Figure \ref{fig:FMR_offset} shows these results. A Kolmogorov-Smirnov test suggests there is no significant difference between the metallicity distributions of galaxies with and without detected absorption (relative to the FMR). Instead, there is a consistent offset of $\approx$0.1 dex. This is well within the margin of error, and may be due to systematic differences between the mass and SFR estimates we obtain through our SED fitting, and those measured by \citet{curti2020}.

\begin{figure*}
\includegraphics[trim={0.01cm 0 0 0},clip, width=\textwidth]{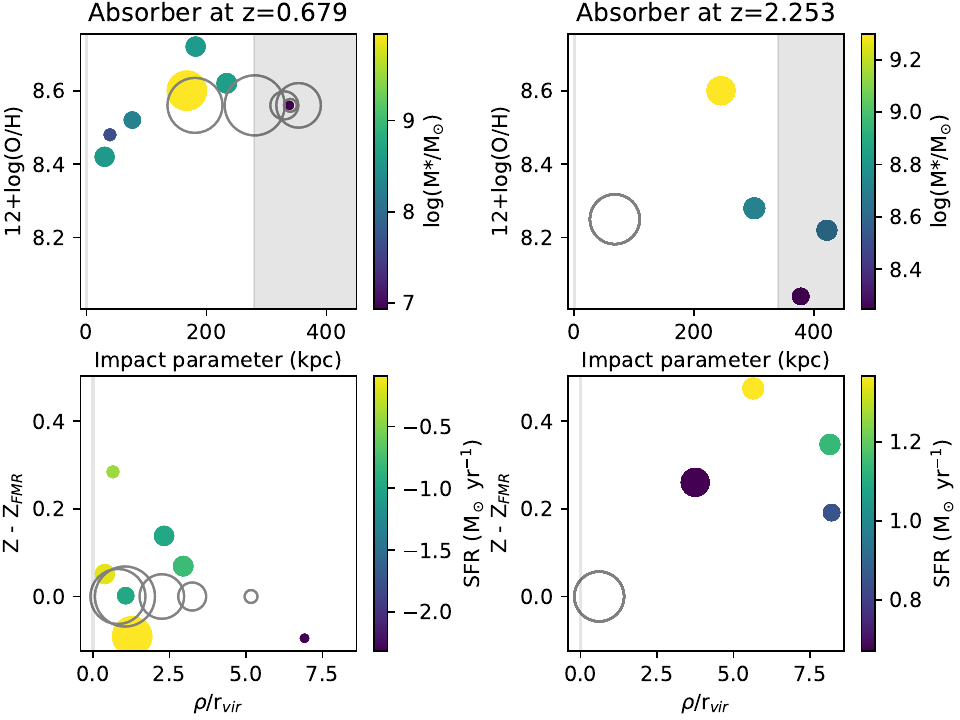}
\caption{The metallicity (upper panels) and metallicity offset from the FMR (lower panels) of galaxies around the two absorption systems for which these measurements are available (within 500 \kms), as a function of impact parameter and $\rho$/r$_{vir}$. Grey circles mark galaxies without a metallicity measurement, which are placed at the median value in the upper panels and at zero in the lower panels by default. Galaxies with measured metallicity are colored by the stellar mass in the upper panels, and the SFR in the lower panels, and sized to indicate stellar mass. The grey shaded regions in the upper panels show the impact parameters beyond which our coverage is incomplete due to the non-symmetry of our MUSE coverage.
\label{fig:metal_groups}}
\end{figure*}

\begin{figure*}
\includegraphics[trim={0.1cm 0 0 0},clip, width=\textwidth]{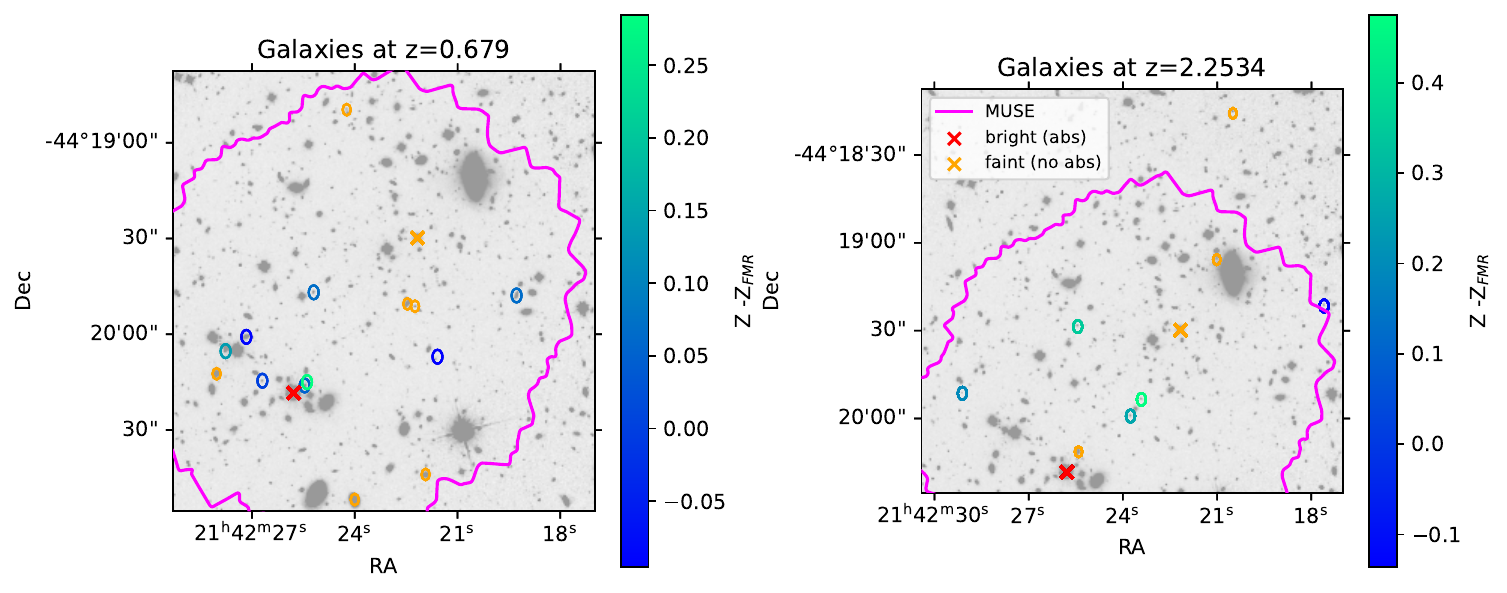}
\caption{The on-sky locations of galaxies around the absorption at z-0.679 (left) and 2.253 (right). The circles showing galaxy locations are colored by metallicity offset from the FMR where this is measured. The red cross marks the location of the absorbing (brighter) quasar. Galaxies without metallicity measurements and the quasar without detected absorption are marked with orange circles and cross respectively. We also show the region covered by MUSE, outside of which our galaxy redshift completeness is substantially lower.
\label{fig:metal_group_layout}}
\end{figure*}

We also consider the galaxies that have measured metallicities as well as nearby absorption. Figure \ref{fig:metal_groups} shows the metallicities of galaxies around the two redshifts which feature galaxies with metallicity measurements within 400 kpc of absorption. These are given as a function of impact parameter to the absorber. The upper panels show the metallicity and impact parameter, whilst the lower panels show the metallicity offset from the FMR and the impact parameter normalized by the galaxy virial radius. 

The higher metallicities seen in the low-redshift group are expected due to evolution in the MZR. Whilst most of the galaxies in the lower-redshift group are scattered around the FMR, those in the high-redshift group lie 0.2 dex or more above the FMR. The low-z group also shows a slight correlation between metallicity and impact parameter, but the metallicity differences are not much larger than the uncertainties on these measurements. However, the low-redshift absorber lies at very small impact parameter to a highly-star-forming galaxy found $\approx$0.5 dex above the FMR, shown by the highest point in the lower-left panel of Figure \ref{fig:metal_groups}.

Mg~II is visible at both redshifts. The equivalent width of the absorber at z=0.679 is comparable to other absorbers in the line-of-sight ($\approx$ 0.9 \AA), whilst that at z=2.253 has a higher EW than any other Mg~II absorber ($\approx$ 2 \AA). 

The on-sky projection of these two groups is shown in Figure \ref{fig:metal_group_layout}. In both cases a large fraction of galaxies do not have metallicity measurements, making it difficult to detect any clear structure in the metallicities that are measured. 

\section{Discussion}\label{sec:discussion}

Using a sample of galaxies and absorbers along two closely-spaced lines-of-sight, we have investigated whether the azimuthal angle distribution of absorption around galaxies or any correlation between absorber properties and galaxy metallicity provides evidence for gas recycling at higher redshifts. This could manifest as a bimodality in the azimuthal angle distribution, indicating separate outflowing and infalling structures similar to those seen around local galaxies \citep[most famously M-82, e.g.][]{bland1988, sorgho2019}, or as a difference in metallicity for galaxies with associated absorption, which could indicate the expulsion or accretion of metal-rich gas from/into the CGM. Although we do not detect this bimodality in the full MUDF sample at a significant level, it is seen in 
a subsample at small impact parameters ($\lesssim$ 2 r$_{vir}$) and around galaxies with high sSFR. This detection of a bimodal distribution is not found to be robust. We do not find a significant difference in metallicity with respect to the FMR between galaxies with and without absorption detected nearby. Here we discuss the implications of these results and their context in the existing literature.

\subsection{Azimuthal Angle Distribution}\label{sec:disc_azimuth}

At low redshifts (z $\lesssim$ 1.5), there are many studies using metal absorption, often Mg~II, that find a bimodal distribution in azimuthal angle \citep[e.g.][]{bouche2012, kacprzak2012, schroetter2019}. However, most of their absorbers lie at small impact parameters (usually $\lesssim 100$ kpc). Creating a sufficiently deep spectroscopic survey that extends to larger radii is observationally expensive, and therefore much more rare. Using three quasar lines-of-sight to improve the efficiency, the survey of Q0107 does extend to much larger scales, and was able to show that at z $<$ 1 this bimodality is evident in H~I and extends to $\approx$ 300 kpc \citep{beckett2021}. As shown in Figure \ref{fig:pa_inc_mudf_all}, our sample also includes galaxy--absorber pairs at impact parameters large enough to detect a bimodality extending out to $\approx$ 300 kpc, although most of our detections are at smaller impact parameters. 

It is also important to note that several studies have found no such result, so these structures are clearly not ubiquitous \citep[e.g.][]{dutta2020, huang2021a}, and their detection is likely to depend on the sample selection. For example, the MEGAFLOW survey is targeted towards strong Mg~II absorption, and therefore small impact parameters in most cases. This leads to less confusion in determining the 'host' galaxy of each absorber, and hence less `noise' in the azimuthal angle distribution, making it easier to detect such a bimodality. The sample splits we apply sometimes reveal a bimodality by removing some of this noise.

For unsaturated absorbers, \citet{lan2017} provide an estimate of H~I column density as a function of Mg~II (2796 \AA) equivalent width. This suggests most of our Mg~II absorbers, which have EWs larger than the point where they begin to saturate (near EWs of $\approx$1 \AA), are associated with column densities of H~I at least 10$^{19}$ cm$^{-2}$. This is far larger than the column densities included in even the high-column-density subsample of the Q0107 data, which used a threshold of 10$^{14}$ cm$^{-2}$.

These strong absorbers are often found within tens of kpc of a galaxy \citep[e.g.][]{rao2011, nielsen2018, dutta2020}. This implies that many of our galaxy absorber pairs at larger impact parameters may not be physically connected, consistent with our observed impact parameter distribution. Such a different population of absorbers would go some way to explaining the differences between the Q0107 and MUDF results described in Section \ref{sec:tests}.


Unfortunately our sample is not large enough to split the 'low-impact-parameter' samples (upper-middle and upper-right panels of Figure \ref{fig:sample_size_mudf}) further, nor can we use lower impact parameter thresholds to limit our sample to the galaxy--absorber pairs most likely to be physically associated. These changes would quickly reduce the sample size to 100 pairs or less, which our resampling shows is too small for a reliable dip-test result. 

When the impact parameters are normalized by the galaxy virial radius a 2$\sigma$ result is found, but splitting the sample at a constant threshold value does not reveal a significant result. This might suggest that the distances to which cool gas can be observed scale with the galaxy mass, as found in groups by \citet{cherrey2024}. Simulations also suggest that the size and number of cool clouds in outflows, as well as the radius at which accreting gas becomes rotationally-supported are expected to evolve with galaxy stellar mass (among other dependencies, \citealt{fielding2022, gronke2022, stern2024}), but detecting such evolution would require a much larger sample than is available for this work.


Aside from impact parameter, we would likely expect the strongest difference to appear in the sample when split by star formation, as stellar feedback is expected to drive most observed outflows around low-mass galaxies. We observe this link between gas distribution and SFR, as shown in Figure \ref{fig:sample_size_mudf}, although it is only significant when considering specific star-formation rate. This could be hidden in the \citet{beckett2021} results, which consist of lower-column-density absobers that are likely to have larger impact parameters. Therefore the time taken for the observed outflow to reach these distances from the galaxy allows the star-formation that drove the outflow to differ from the SFR observed in the galaxy (the travel time for a 200 \kms{ }outflow to reach 300 kpc distance is $\approx$1.5 Gyr).

Any bimodality in the distribution of absorption around low-mass galaxies is very tentative, but our results are consistent with \citet{beckett2021}. The metal content of the z$\sim$3 IGM is found to require enrichment from low-mass galaxies due (in part) to their higher abundance \citep[e.g.][]{booth2012}, which would lead us to expect a bimodal distribution around low-stellar-mass galaxies.


The results obtained when splitting the sample by inclination are not significant, but match the Q0107 results well. However, there is a notable difference in redshift, with the Q0107 data showing a strong bimodality at z $<$ 0.5, whereas any redshift evolution in the MUDF suggests a stronger result at higher redshifts (z $>$ 1.2). The trend in our MUDF data may be due to galaxy selection, where galaxies without emission lines cannot easily be detected at high-redshifts, leaving a more highly SFR-biased sample. 

A comparison of galaxy--H~I absorber pairs at high and low redshifts using the Q0107 and MUDF datasets will require further observations to identify galaxies at z $>$ 2, and hence eliminate some possible observational effects that could be masking any redshift evolution in the CGM.

\subsection{Metallicity Dependence on Absorption}\label{sec:disc_metals}

No firm conclusion has yet been reached as to whether absorber metallicity varies as a function of azimuthal angle or galaxy metallicity. Some studies using data including H~I and metal absorption around isolated galaxies at low redshifts (z $\lesssim$ 1) have not shown strong evidence of these relationships \citep{kacprzak2019, pointon2019, weng2023a}, but \citet{wendt2021} do show evidence for more metal-enriched gas closer to the galaxy minor axis. Our lack of H~I coverage (at the same redshifts where we can measure galaxy metallicities) means that we cannot currently measure CGM metallicities and compare our results with these studies; we would require a larger sample of galaxies at z $\gtrsim$ 2.5 in order to construct a comparable sample and search for redshift evolution in the metallicity of the CGM. Although Zn, Si and S lines can be used to estimate dust depletion, a useful proxy for metallicity \citep[e.g.][]{decia2018}, these lines are only detected for a small fraction of our absorbers. We can therefore only compare absorber EW and galaxy metallicity.

Our measurements of absorber properties and comparison with galaxy metallicity failed to reveal any correlation between absorber EW and galaxy metallicity, or any difference in the metallicities of galaxies with and without absorption. This is at least in part due to the small sample size, as the redshift ranges for which absorption is detected and those in which we can measure the emission lines required to calculate a metallicity only partially overlap. The metallicity measurements also require strong line detections to accurately measure flux, so can only be done on a fraction of the galaxies even in these redshift ranges.

The fundamental metallicity relation (FMR), against which we compare, includes a dependence on SFR, and therefore encodes a reduction in metallicity from the accretion of pristine gas and the expulsion of metal-enriched gas due to stellar (and AGN) feedback \citep[e.g][]{greener2022, yang2022}. Our comparison of absorber properties and galaxy metallicity offset from the FMR is therefore not testing for the presence of feedback and recycling, but for whether the detection of absorption indicates higher-than-expected levels of recycling.

As discussed above, we find only two absorbers within 400 kpc of galaxies with metallicity measurements. However, the closest galaxy at z=2.25 has no metallicity measurement (as shown in Figures \ref{fig:metal_groups} and \ref{fig:metal_group_layout}) due to a lack of clear emission lines. Only the two closest galaxies at z=0.679 are probed within their virial radius and have measured metallicities. This is far too small a sample to make any strong conclusions, although it is interesting that one of these galaxies has a far higher metallicity than expected from the FMR.

In \citet{revalski2024}, a tentative enhancement in metallicity among group galaxies is found ($\approx$0.1 dex at z $\lesssim$ 1 and 0.2 dex at z $\gtrsim$ 1), although is not significant and the high-z result is driven in large part by the group at z=2.25 that we show in Figure \ref{fig:metal_groups}. All four group galaxies at z=2.25 have higher metallicity than predicted by the FMR. However, the absorption, if it can be traced to a single galaxy, is most likely associated with the high-mass galaxy closest to the line-of-sight, for which we do not have a metallicity measurement.

\section{Summary and Conclusions} \label{sec:conclusions}

In this work we have utilized the MUSE Ultra Deep Field, one of the deepest IFU surveys observed to-date of a field including two bright quasars, to investigate the connection between metal absorption lines in the background quasar spectra and the number and properties of associated galaxies. Our findings are as follows:

\begin{enumerate}
    \item A positive correlation (2.5 $\sigma$) between absorber equivalent width (EW) and number of nearby galaxies is found for Fe~II absorption systems in these sightlines. We do not find the expected anti-correlation with impact parameter, but are limited to a small sample of Fe~II absorbers due to saturation in the Mg~II sample.

    \item We do not detect a bimodality in the azimuthal angle distribution of galaxy--absorber pairs, in contrast with the lower-redshift data from \citet{beckett2021}. When we restrict the sample to those with impact parameters $<$ 2 r$_{vir}$, we do then find a bimodality at the  2$\sigma$ level of significance. This would suggest metal absorption at 0.5 $<$ z $<$ 3 does trace disk and outflowing structures, but to smaller scales than H~I absorption at lower redshifts.

    \item A bimodality is found (at $\approx$ 2$\sigma$ significance) in the azimuthal angles of absorbers around our galaxy samples with higher-than-average specific star-formation rate. The samples with a stellar masses less than 10$^{9.5}$ M$_{\odot}$ and with higher-than-average star-formation rate also show some apparent bimodality, but do not reach the 2$\sigma$ level of significance. These indicate that, where these disk and outflow structures exist, they are likely linked to star formation in galaxies, with larger extents possible in shallower gravitational potentials.

    \item We find no evidence of any correlation between galaxy metallicity (as offset from the fundamental metallicity relation) and absorber EW. This consistency with the FMR would suggest that the mechanisms leading to observed absorption are already encoded into our parameterization of the FMR.

    \item Only two galaxies with metallicity measurements have detected absorption within 500 \kms{ } along the line-of-sight and within their virial radius. One of these has a high metallicity (0.3 dex above the FMR) and high SFR (above the main sequence), indicating that this absorption system could be related to processes not encoded in the FMR.
\end{enumerate}

The results we have obtained in this work are necessarily tentative because of the limited sample size. This is partly due to the misalignment between the redshift ranges in which galaxy and absorber properties can be measured using this dataset (see Figure \ref{fig:z_dists}). For example, above z $\approx$ 2.4, the only strong emission line we can detect is [O~II], but this is generally hard to distinguish from Ly$\alpha$ in the low-resolution HST grism spectra, so we do not have a confident redshift for most galaxies in this range. Unfortunately, this is also the only redshift range for which we have coverage of H~I absorption, which would enable us to measure absorber metallicities. 

JWST observations of this field, in particular the extremely large wavelength range covered by the NIRSpec prism, would allow galaxy redshifts and metallicities to be measured across this range and allow us to fully exploit this unique dataset. We would increase the number of galaxy--absorber pairs with galaxy metallicities by a factor of $\approx$4. This would allow direct comparison of galaxy and absorber metallicities for tens of galaxy--absorber pairs, similar to \citet{kacprzak2019} but at redshift z$\approx$3. Some of these pairs would also feature absorption in both quasar sightlines, allowing deeper analysis of absorber coherence lengths and the metallicity distribution around galaxies.

Currently, given the relatively rare bright background sources required for these observations,
and the depth required to obtain high signal-to-noise emission line measurements from high-redshift galaxies, surveys for which we can test the link between galaxies and absorbing gas at z $>$ 2 are rare. However, absorption remains our most powerful method of detecting low-column-density material at large impact parameters around distant galaxies. With only a pencil-beam view (or in a few cases, a small number of pencil-beams) of the gas around any individual galaxy, we will need to continue building large samples of galaxies and absorbers in order to constrain the role of gas feedback and recycling in our models of galaxy evolution.

\section*{acknowledgments}

We thank the anonymous reviewer for their thoughtful comments that have improved this manuscript. This work is based on observations with the NASA/ESA Hubble Space Telescope obtained from the MAST Data Archive at the Space Telescope Science Institute, which is operated by the Association of Universities for Research in Astronomy, Incorporated, under NASA contract NAS5-26555. Support for program numbers 15637 and 15968 was provided through a grant from STScI under NASA contract NAS5-26555. These observations are associated with program numbers 6631, 15637, and 15968. The VLT portion of this project has received funding from the European Research Council (ERC) under the European Union’s Horizon 2020 research and innovation programme (grant agreement No 757535) and by Fondazione Cariplo (grant No 2018-2329). PD acknowledges support from the NWO grant 016.VIDI.189.162 (``ODIN") and from the European Commission's and University of Groningen's CO-FUND Rosalind Franklin program. SC gratefully acknowledges support from the European Research Council (ERC) under the European Union’s Horizon 2020 Research and Innovation programme grant agreement No 864361. This research has made use of NASA's Astrophysics Data System.



%

\vspace{5mm}
\facilities{HST(WFC3, WFPC2), VLT(MUSE, UVES, HAWK-I)}


\software{
Astropy \citep{astropycollaboration2013, astropycollaboration2018, astropycollaboration2022},
Galfit \citep{peng2002},
IPython \citep{perez2007},
Jupyter \citep{kluyver2016},
Matplotlib \citep{hunter2007, caswell2023}
Numpy \citep{harris2020},
Scipy \citep{virtanen2020},
Source Extractor \citep{1996A&AS..117..393B},
Statmorph \citep{rodriguez-gomez2019}
          }



\appendix

\section{Position angle measurements} \label{sec:pa_measurements}

If the systematic uncertainties on our galaxy orientation measurements are underestimated, it is possible that any bimodality could be `smeared-out' by these errors. In order to check this, we run both the GALFIT \citep{peng2002, peng2010a} and STATMORPH \citep{rodriguez-gomez2019} software on this dataset and compare the resulting position angle estimates. 

GALFIT models the light profile of a galaxy, enabling decomposition into bulge and disk components, and fits these through chi-squared minimization. However, given the large number of galaxies to model, we use only a single Sersic profile. We can confirm that the resulting fits account for most of the light from each galaxy in the images by inspecting the residuals, and extract parameters related to the galaxy orientation, namely the position angle and inclination.

\begin{figure}
\includegraphics[trim={0.001cm 0 0 0},clip, width=\columnwidth]{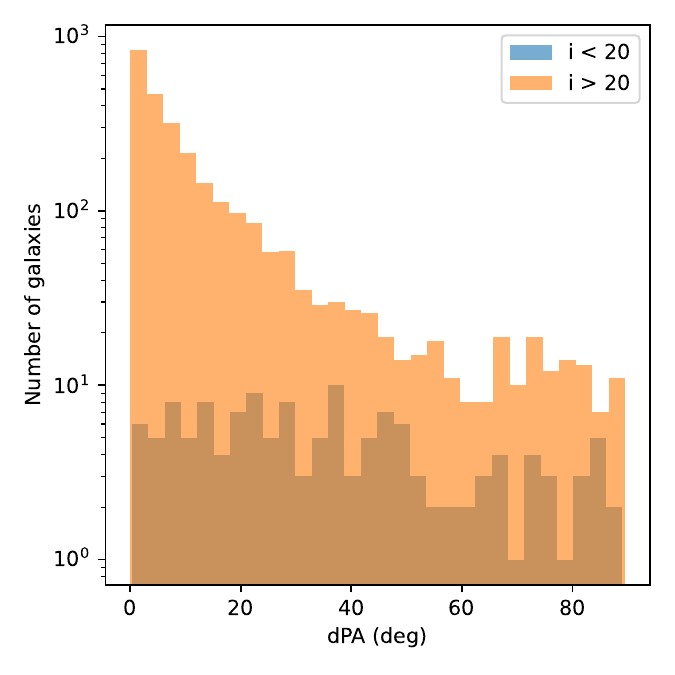}
\caption{The difference in position angle estimates through GALFIT and \textsc{statmorph}, split into two samples with inclinations above and below $20^{\circ}$.
\label{fig:PA_inc_bins}}
\end{figure}

\begin{figure}
\includegraphics[trim={0.0001cm 0 0 0},clip, width=\columnwidth]{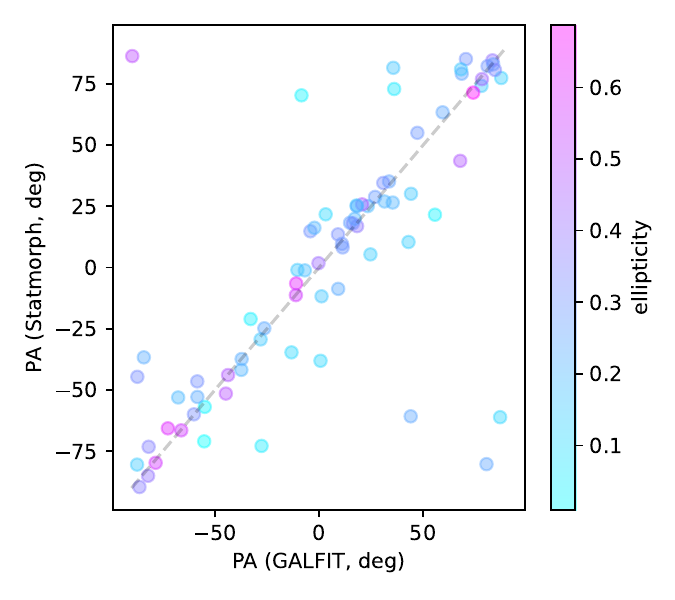}
\caption{The position angles of galaxies estimated through \textsc{GALFIT} and through \textsc{statmorph}. Points are coloured by ellipticity measured using \textsc{GALFIT}. Only galaxies with well-determined redshifts within 500 \kms{ }of detected absorption are shown. The black dashed line indicates identical position angles between the two codes.
\label{fig:detGFvsSM}}
\end{figure}

Figure \ref{fig:PA_inc_bins} compares the resulting position angle measurements with those from \textsc{statmorph} described in Section \ref{sec:data}. Most galaxies clearly have very similar position angles in the two measurements, confirming that our Statmorph orientations are reliable. We obtain a 1$\sigma$ difference of $12^{\circ}$. We also split the sample by inclination at $20^{\circ}$. These results suggest a 1$\sigma$ uncertainty of $10^{\circ}$ for galaxies with i $>$ $20^{\circ}$ (those that are not close to face-on), and a much larger uncertainty of $40^{\circ}$ for those closer to face-on.

Any position-angle dependency in the CGM will be most clear for galaxies close to edge-on, where the projected major and minor axes most closely represent directions parallel and perpendicular to the plane of any galaxy disk. \citet{beckett2021} show a bimodality using bins of width $30^{\circ}$, and have typical uncertainties of $\approx 15^{\circ}$ so uncertainties of less than $10^{\circ}$ should not substantially affect the detection of such a result. Errors in our position angle measurements are therefore unlikely to contribute to the lack of bimodality seen in our results.

We also manually check for any systematic differences by looking specifically at galaxies with large differences between their position angles as measured using GALFIT and Statmorph. 27 galaxies have ellipticities greater than 0.5 and PA measurements that differ by more than 20$^{\circ}$. Of these, several appear to have an extremely low surface brightness that would make it difficult to determine a PA, others lie very close to other objects (so are likely to have their fits affected by the companion), and others appear to have multiple components with different position angles (possibly other galaxies with small projected separations, or structure in the galaxy being measured). 

Where the Statmorph and GALFIT results differ due to multiple objects or structures within objects, Statmorph tends to flag these as `suspect' as the flux distribution is not smooth, but will often combine the sources into a single ellipse, whereas GALFIT will fit a single profile, and therefore select one object or structure. A test `by-eye' to see which of the two estimates is a better fit overall does not prefer one of these methods over the other. In some cases, the 'combined' profile used by Statmorph contains a single source, such as a spiral galaxy, where it provides a better estimate than GALFIT, which often struggles to model this and may only fit to a bulge- or bar-like central structure. This approximately balances cases where GALFIT is preferred due to fitting a single galaxy whereas Statmorph combines multiple sources. 

We note that a large proportion (2145 of 3186) of the Statmorph sources are flagged as `suspect', but most of these are still measured successfully, with a 1$\sigma$ uncertainty of 15$^{\circ}$.

We compare the two measures for the galaxies within 500 \kms{ }of detected absorption, showing the results in Figure \ref{fig:detGFvsSM}. 13 of these galaxies have measured PAs differing by more than 20$^{\circ}$ (of 74 total galaxies near absorber redshifts). Visual inspection of these galaxies reveals a similar trend, with neither code preferred overall. 7 of these 13 are close to face-on, with the other 6 either compound or complex objects generally following the trend discussed above. 

This behaviour can be tuned using the threshold options available within SExtractor, and masking options within GALFIT, but given the small number of galaxies affected by our choice of position angle measurement, we do not attempt to further adjust the parameters. For most galaxies we use the Statmorph values given in the MUDF data release for this imaging \citep{revalski2023}, whilst switching to the GALFIT results for the individual galaxies for which GALFIT is clearly preferred.


\bibliography{zotero_full_240807, sample631}{}

\begin{thebibliography}{}
\expandafter\ifx\csname natexlab\endcsname\relax\def\natexlab#1{#1}\fi
\providecommand{\url}[1]{\href{#1}{#1}}
\providecommand{\dodoi}[1]{doi:~\href{http://doi.org/#1}{\nolinkurl{#1}}}
\providecommand{\doeprint}[1]{\href{http://ascl.net/#1}{\nolinkurl{http://ascl.net/#1}}}
\providecommand{\doarXiv}[1]{\href{https://arxiv.org/abs/#1}{\nolinkurl{https://arxiv.org/abs/#1}}}

\bibitem[{Akaike(1974)}]{akaike1974}
Akaike, H. 1974, IEEE Transactions on Automatic Control, 19, 716

\bibitem[{Andrews \& Martini(2013)}]{andrews2013}
Andrews, B.~H., \& Martini, P. 2013, ApJ, 765, 140

\bibitem[{{Astropy Collaboration} {et~al.}(2013){Astropy Collaboration},
  Robitaille, Tollerud, Greenfield, Droettboom, Bray, Aldcroft, Davis,
  Ginsburg, {Price-Whelan}, Kerzendorf, Conley, Crighton, Barbary, Muna,
  Ferguson, Grollier, Parikh, Nair, Unther, Deil, Woillez, Conseil, Kramer,
  Turner, Singer, Fox, Weaver, Zabalza, Edwards, Azalee~Bostroem, Burke, Casey,
  Crawford, Dencheva, Ely, Jenness, Labrie, Lim, Pierfederici, Pontzen, Ptak,
  Refsdal, Servillat, \& Streicher}]{astropycollaboration2013}
{Astropy Collaboration}, Robitaille, T.~P., Tollerud, E.~J., {et~al.} 2013,
  A\&A, 558, A33

\bibitem[{{Astropy Collaboration} {et~al.}(2018){Astropy Collaboration},
  {Price-Whelan}, Sip{\H o}cz, G{\"u}nther, Lim, Crawford, Conseil, Shupe,
  Craig, Dencheva, Ginsburg, VanderPlas, Bradley, {P{\'e}rez-Su{\'a}rez}, {de
  Val-Borro}, Aldcroft, Cruz, Robitaille, Tollerud, Ardelean, Babej, Bach,
  Bachetti, Bakanov, Bamford, Barentsen, Barmby, Baumbach, Berry, Biscani,
  Boquien, Bostroem, Bouma, Brammer, Bray, Breytenbach, Buddelmeijer, Burke,
  Calderone, Cano~Rodr{\'i}guez, Cara, Cardoso, Cheedella, Copin, Corrales,
  Crichton, D'Avella, Deil, Depagne, Dietrich, Donath, Droettboom, Earl, Erben,
  Fabbro, Ferreira, Finethy, Fox, Garrison, Gibbons, Goldstein, Gommers, Greco,
  Greenfield, Groener, Grollier, Hagen, Hirst, Homeier, Horton, Hosseinzadeh,
  Hu, Hunkeler, Ivezi{\'c}, Jain, Jenness, Kanarek, Kendrew, Kern, Kerzendorf,
  Khvalko, King, Kirkby, Kulkarni, Kumar, Lee, Lenz, Littlefair, Ma, Macleod,
  Mastropietro, McCully, Montagnac, Morris, Mueller, Mumford, Muna, Murphy,
  Nelson, Nguyen, Ninan, N{\"o}the, Ogaz, Oh, Parejko, Parley, Pascual, Patil,
  Patil, Plunkett, Prochaska, Rastogi, Reddy~Janga, Sabater, Sakurikar,
  Seifert, Sherbert, {Sherwood-Taylor}, Shih, Sick, Silbiger, Singanamalla,
  Singer, Sladen, Sooley, Sornarajah, Streicher, Teuben, Thomas, Tremblay,
  Turner, Terr{\'o}n, {van Kerkwijk}, {de la Vega}, Watkins, Weaver, Whitmore,
  Woillez, Zabalza, \& {Astropy Contributors}}]{astropycollaboration2018}
{Astropy Collaboration}, {Price-Whelan}, A.~M., Sip{\H o}cz, B.~M., {et~al.}
  2018, AJ, 156, 123

\bibitem[{{Astropy Collaboration} {et~al.}(2022){Astropy Collaboration},
  {Price-Whelan}, Lim, Earl, Starkman, Bradley, Shupe, Patil, Corrales,
  Brasseur, N{\"o}the, Donath, Tollerud, Morris, Ginsburg, Vaher, Weaver,
  Tocknell, Jamieson, {van Kerkwijk}, Robitaille, Merry, Bachetti, G{\"u}nther,
  Aldcroft, {Alvarado-Montes}, Archibald, B{\'o}di, Bapat, Barentsen,
  Baz{\'a}n, Biswas, Boquien, Burke, Cara, Cara, Conroy, Conseil, Craig, Cross,
  Cruz, D'Eugenio, Dencheva, Devillepoix, Dietrich, Eigenbrot, Erben, Ferreira,
  {Foreman-Mackey}, Fox, Freij, Garg, Geda, Glattly, Gondhalekar, Gordon,
  Grant, Greenfield, Groener, Guest, Gurovich, Handberg, Hart,
  {Hatfield-Dodds}, Homeier, Hosseinzadeh, Jenness, Jones, Joseph, Kalmbach,
  Karamehmetoglu, Ka{\l}uszy{\'n}ski, Kelley, Kern, Kerzendorf, Koch, Kulumani,
  Lee, Ly, Ma, MacBride, Maljaars, Muna, Murphy, Norman, O'Steen, Oman,
  Pacifici, Pascual, {Pascual-Granado}, Patil, Perren, Pickering, Rastogi,
  Roulston, Ryan, Rykoff, Sabater, Sakurikar, Salgado, Sanghi, Saunders,
  Savchenko, Schwardt, {Seifert-Eckert}, Shih, Jain, Shukla, Sick, Simpson,
  Singanamalla, Singer, Singhal, Sinha, Sip{\H o}cz, Spitler, Stansby,
  Streicher, {\v S}umak, Swinbank, Taranu, Tewary, Tremblay, {de Val-Borro},
  Van~Kooten, Vasovi{\'c}, Verma, {de Miranda Cardoso}, Williams, Wilson,
  Winkel, {Wood-Vasey}, Xue, Yoachim, Zhang, Zonca, \& {Astropy Project
  Contributors}}]{astropycollaboration2022}
{Astropy Collaboration}, {Price-Whelan}, A.~M., Lim, P.~L., {et~al.} 2022, ApJ,
  935, 167

\bibitem[{Bacon {et~al.}(2010)Bacon, Accardo, Adjali, Anwand, Bauer, Biswas,
  Blaizot, Boudon, {Brau-Nogue}, Brinchmann, Caillier, Capoani, Carollo,
  Contini, Couderc, Daguis{\'e}, Deiries, Delabre, Dreizler, Dubois, Dupieux,
  Dupuy, Emsellem, Fechner, Fleischmann, Fran{\c c}ois, Gallou, Gharsa,
  Glindemann, Gojak, Guiderdoni, Hansali, Hahn, Jarno, Kelz, Koehler,
  Kosmalski, Laurent, Le~Floch, Lilly, Lizon, Loupias, Manescau, Monstein,
  Nicklas, Olaya, Pares, Pasquini, {P{\'e}contal-Rousset}, Pell{\'o}, Petit,
  Popow, Reiss, Remillieux, Renault, Roth, Rupprecht, Serre, Schaye, Soucail,
  Steinmetz, Streicher, Stuik, Valentin, Vernet, Weilbacher, Wisotzki, \&
  Yerle}]{bacon2010}
Bacon, R., Accardo, M., Adjali, L., {et~al.} 2010, SPIE, 7735, 773508

\bibitem[{Bacon {et~al.}(2017)Bacon, Conseil, Mary, Brinchmann, Shepherd,
  Akhlaghi, Weilbacher, Piqueras, Wisotzki, Lagattuta, Epinat, Guerou, Inami,
  Cantalupo, Courbot, Contini, Richard, Maseda, Bouwens, Bouche, Kollatschny,
  Schaye, Marino, Pello, Herenz, Guiderdoni, \& Carollo}]{bacon2017}
Bacon, R., Conseil, S., Mary, D., {et~al.} 2017, A\&A, 608, A1.
\newblock \doarXiv{1710.03002}

\bibitem[{Bacon {et~al.}(2023)Bacon, Brinchmann, Conseil, Maseda, Nanayakkara,
  Wendt, Bacher, Mary, Weilbacher, Krajnovi{\'c}, Boogaard, Bouch{\'e},
  Contini, Epinat, Feltre, Guo, Herenz, Kollatschny, Kusakabe, Leclercq,
  {Michel-Dansac}, Pello, Richard, Roth, Salvignol, Schaye, Steinmetz, Tresse,
  Urrutia, Verhamme, Vitte, Wisotzki, \& Zoutendijk}]{bacon2023}
Bacon, R., Brinchmann, J., Conseil, S., {et~al.} 2023, A\&A, 670, A4

\bibitem[{Beckett {et~al.}(2021)Beckett, Morris, Fumagalli, Bielby, Tejos,
  Schaye, Jannuzi, \& Cantalupo}]{beckett2021}
Beckett, A., Morris, S.~L., Fumagalli, M., {et~al.} 2021, MNRAS, 506, 2574

\bibitem[{Behroozi {et~al.}(2010)Behroozi, Conroy, \& Wechsler}]{behroozi2010}
Behroozi, P.~S., Conroy, C., \& Wechsler, R.~H. 2010, ApJ, 717, 379.
\newblock \doarXiv{1001.0015}

\bibitem[{Bertin \& Arnouts(1996)}]{bertin1996}
Bertin, E., \& Arnouts, S. 1996, ApJS, 117, 393

\bibitem[{{Bertin} \& {Arnouts}(1996)}]{1996A&AS..117..393B}
{Bertin}, E., \& {Arnouts}, S. 1996, \aaps, 117, 393,
  \dodoi{10.1051/aas:1996164}

\bibitem[{Bielby {et~al.}(2019)Bielby, Stott, Cullen, Tripp, Burchett,
  Fumagalli, Morris, Tejos, Crain, Bower, \& Prochaska}]{bielby2019}
Bielby, R.~M., Stott, J.~P., Cullen, F., {et~al.} 2019, MNRAS, 486, 21.
\newblock \doarXiv{1809.05544}

\bibitem[{Bland \& Tully(1988)}]{bland1988}
Bland, J., \& Tully, B. 1988, Nature, 334, 43

\bibitem[{Booth {et~al.}(2012)Booth, Schaye, Delgado, \&
  Dalla~Vecchia}]{booth2012}
Booth, C.~M., Schaye, J., Delgado, J.~D., \& Dalla~Vecchia, C. 2012, MNRAS,
  420, 1053

\bibitem[{Bordoloi {et~al.}(2011)Bordoloi, Lilly, Knobel, Bolzonella, Kampczyk,
  Carollo, Iovino, Zucca, Contini, Kneib, Le~Fevre, Mainieri, Renzini,
  Scodeggio, Zamorani, Balestra, Bardelli, Bongiorno, Caputi, Cucciati, {de la
  Torre}, {de Ravel}, Garilli, Kova{\v c}, Lamareille, Le~Borgne, Le~Brun,
  Maier, Mignoli, Pello, Peng, Perez~Montero, Presotto, Scarlata, Silverman,
  Tanaka, Tasca, Tresse, Vergani, Barnes, Cappi, Cimatti, Coppa, Diener,
  Franzetti, Koekemoer, {L{\'o}pez-Sanjuan}, McCracken, Moresco, Nair, Oesch,
  Pozzetti, \& Welikala}]{bordoloi2011}
Bordoloi, R., Lilly, S.~J., Knobel, C., {et~al.} 2011, ApJ, 743, 10

\bibitem[{Bouch{\'e} {et~al.}(2012)Bouch{\'e}, Hohensee, Vargas, Kacprzak,
  Martin, Cooke, \& Churchill}]{bouche2012}
Bouch{\'e}, N., Hohensee, W., Vargas, R., {et~al.} 2012, MNRAS, 426, 801

\bibitem[{Bruzual \& Charlot(2003)}]{bruzual2003}
Bruzual, G., \& Charlot, S. 2003, MNRAS, 344, 1000

\bibitem[{Burchett {et~al.}(2021)Burchett, Rubin, Prochaska, Coil,
  Rickards~Vaught, \& Hennawi}]{burchett2021}
Burchett, J.~N., Rubin, K. H.~R., Prochaska, J.~X., {et~al.} 2021, ApJ, 909,
  151.
\newblock \doarXiv{2005.03017}

\bibitem[{Byler {et~al.}(2017)Byler, Dalcanton, Conroy, \& Johnson}]{byler2017}
Byler, N., Dalcanton, J.~J., Conroy, C., \& Johnson, B.~D. 2017, ApJ, 840, 44

\bibitem[{Calzetti {et~al.}(2000)Calzetti, Armus, Bohlin, Kinney, Koornneef, \&
  {Storchi-Bergmann}}]{calzetti2000}
Calzetti, D., Armus, L., Bohlin, R.~C., {et~al.} 2000, ApJ, 533, 682.
\newblock \doarXiv{astro-ph/9911459}

\bibitem[{Caswell {et~al.}(2023)Caswell, {Sales de Andrade}, Lee, Droettboom,
  Hoffmann, Klymak, Hunter, Firing, Stansby, Varoquaux, Hedegaard~Nielsen,
  Gustafsson, Sunden, Root, May, Elson, Sepp{\"a}nen, {hannah}, Lee, Dale,
  McDougall, Straw, Hobson, Lucas, Comer, Gohlke, Vincent, Yu, Ma, \&
  Silvester}]{caswell2023}
Caswell, T.~A., {Sales de Andrade}, E., Lee, A., {et~al.} 2023,
  Matplotlib/Matplotlib: {{REL}}: V3.7.4

\bibitem[{Chabrier(2003)}]{chabrier2003}
Chabrier, G. 2003, PASP, 115, 763

\bibitem[{Chen {et~al.}(2021)Chen, Steidel, Erb, Law, Trainor, Reddy, Shapley,
  Pahl, Strom, Lamb, Li, \& Rudie}]{chen2021a}
Chen, Y., Steidel, C.~C., Erb, D.~K., {et~al.} 2021, MNRAS, 508, 19

\bibitem[{Cherrey {et~al.}(2024)Cherrey, Bouch{\'e}, Zabl, Schroetter, Wendt,
  Langan, Richard, Schaye, Mercier, Epinat, \& Contini}]{cherrey2024}
Cherrey, M., Bouch{\'e}, N.~F., Zabl, J., {et~al.} 2024, MNRAS, 528, 481

\bibitem[{Conroy(2013)}]{conroy2013}
Conroy, C. 2013, ARA\&A, 51, 393

\bibitem[{Cupani {et~al.}(2016)Cupani, D'Odorico, Cristiani,
  {Gonz{\'a}lez-Hern{\'a}ndez}, Lovis, Sousa, Calderone, Cirami, Marcantonio,
  \& M{\'e}gevand}]{cupani2016}
Cupani, G., D'Odorico, V., Cristiani, S., {et~al.} 2016, in Software and
  {{Cyberinfrastructure}} for {{Astronomy IV}}, Vol. 9913 (SPIE), 717--732

\bibitem[{Curti {et~al.}(2017)Curti, Cresci, Mannucci, Marconi, Maiolino, \&
  Esposito}]{curti2017}
Curti, M., Cresci, G., Mannucci, F., {et~al.} 2017, MNRAS, 465, 1384

\bibitem[{Curti {et~al.}(2020)Curti, Mannucci, Cresci, \& Maiolino}]{curti2020}
Curti, M., Mannucci, F., Cresci, G., \& Maiolino, R. 2020, MNRAS, 491, 944

\bibitem[{Curti {et~al.}(2024)Curti, Maiolino, {Curtis-Lake}, Chevallard,
  Carniani, D'Eugenio, Looser, Scholtz, Charlot, Cameron, {\"U}bler, Witstok,
  Boyett, Laseter, Sandles, Arribas, Bunker, Giardino, Maseda, Rawle,
  Rodr{\'i}guez Del~Pino, Smit, Willott, Eisenstein, Hausen, Johnson, Rieke,
  Robertson, Tacchella, Williams, Willmer, Baker, Bhatawdekar, Egami, Helton,
  Ji, Kumari, Perna, Shivaei, \& Sun}]{curti2024}
Curti, M., Maiolino, R., {Curtis-Lake}, E., {et~al.} 2024, A\&A, 684, A75

\bibitem[{Davies {et~al.}(2021)Davies, Crain, \& Pontzen}]{davies2021a}
Davies, J.~J., Crain, R.~A., \& Pontzen, A. 2021, MNRAS, 501, 236.
\newblock \doarXiv{2006.13221}

\bibitem[{Dayal {et~al.}(2013)Dayal, Ferrara, \& Dunlop}]{dayal2013}
Dayal, P., Ferrara, A., \& Dunlop, J.~S. 2013, MNRAS, 430, 2891

\bibitem[{De~Cia {et~al.}(2018)De~Cia, Ledoux, Petitjean, \&
  Savaglio}]{decia2018}
De~Cia, A., Ledoux, C., Petitjean, P., \& Savaglio, S. 2018, A\&A, 611, A76

\bibitem[{DeFelippis {et~al.}(2020)DeFelippis, Genel, Bryan, Nelson, Pillepich,
  \& Hernquist}]{defelippis2020}
DeFelippis, D., Genel, S., Bryan, G.~L., {et~al.} 2020, ApJ, 895, 17.
\newblock \doarXiv{2004.07846}

\bibitem[{Dekker {et~al.}(2000)Dekker, D'Odorico, Kaufer, Delabre, \&
  Kotzlowski}]{dekker2000}
Dekker, H., D'Odorico, S., Kaufer, A., Delabre, B., \& Kotzlowski, H. 2000,
  SPIE, 4008, 534

\bibitem[{Dutta {et~al.}(2020)Dutta, Fumagalli, Fossati, Lofthouse, Prochaska,
  Arrigoni~Battaia, Bielby, Cantalupo, Cooke, Murphy, \& O'Meara}]{dutta2020}
Dutta, R., Fumagalli, M., Fossati, M., {et~al.} 2020, MNRAS, 499, 5022

\bibitem[{Dutta {et~al.}(2021)Dutta, Fumagalli, Fossati, Bielby, Stott,
  Lofthouse, Cantalupo, Cullen, Crain, Tripp, Prochaska, Arrigoni~Battaia,
  Burchett, Fynbo, Murphy, Schaye, Tejos, \& Theuns}]{dutta2021}
---. 2021, MNRAS, 508, 4573

\bibitem[{Dutta {et~al.}(2023)Dutta, Fossati, Fumagalli, Revalski, Lofthouse,
  Nelson, Papini, Rafelski, Cantalupo, Arrigoni~Battaia, Dayal, Longobardi,
  P{\'e}roux, Prichard, \& Prochaska}]{dutta2023a}
Dutta, R., Fossati, M., Fumagalli, M., {et~al.} 2023, MNRAS, 522, 535

\bibitem[{Fielding \& Bryan(2022)}]{fielding2022}
Fielding, D.~B., \& Bryan, G.~L. 2022, ApJ, 924, 82

\bibitem[{Fossati {et~al.}(2019{\natexlab{a}})Fossati, Fumagalli, Gavazzi,
  Consolandi, Boselli, Yagi, Sun, \& Wilman}]{fossati2019a}
Fossati, M., Fumagalli, M., Gavazzi, G., {et~al.} 2019{\natexlab{a}}, MNRAS,
  484, 2212

\bibitem[{Fossati {et~al.}(2018)Fossati, Mendel, Boselli, Cuillandre, Vollmer,
  Boissier, Consolandi, Ferrarese, Gwyn, Amram, Boquien, Buat, Burgarella,
  Cortese, C{\^o}t{\'e}, C{\^o}t{\'e}, Durrell, Fumagalli, Gavazzi,
  {Gomez-Lopez}, Hensler, Koribalski, Longobardi, Peng, Roediger, Sun, \&
  Toloba}]{fossati2018}
Fossati, M., Mendel, J.~T., Boselli, A., {et~al.} 2018, A\&A, 614, A57

\bibitem[{Fossati {et~al.}(2019{\natexlab{b}})Fossati, Fumagalli, Lofthouse,
  D'Odorico, Lusso, Cantalupo, Cooke, Cristiani, Haardt, Morris, Peroux,
  Prichard, Rafelski, Smail, \& Theuns}]{fossati2019}
Fossati, M., Fumagalli, M., Lofthouse, E.~K., {et~al.} 2019{\natexlab{b}},
  MNRAS, 490, 1451.
\newblock \doarXiv{1909.04672}

\bibitem[{French \& Wakker(2020)}]{french2020}
French, D.~M., \& Wakker, B.~P. 2020, ApJ, 897, 151.
\newblock \doarXiv{2006.09323}

\bibitem[{Galbiati {et~al.}(2023)Galbiati, Fumagalli, Fossati, Lofthouse,
  Dutta, Prochaska, Murphy, \& Cantalupo}]{galbiati2023}
Galbiati, M., Fumagalli, M., Fossati, M., {et~al.} 2023, MNRAS, 524, 3474

\bibitem[{Garling {et~al.}(2024)Garling, Peter, Spekkens, Sand, Hargis,
  Crnojevi{\'c}, \& Carlin}]{garling2024}
Garling, C.~T., Peter, A. H.~G., Spekkens, K., {et~al.} 2024, MNRAS, 528, 365

\bibitem[{Gnat \& Sternberg(2007)}]{gnat2007}
Gnat, O., \& Sternberg, A. 2007, ApJS, 168, 213

\bibitem[{Greener {et~al.}(2022)Greener, {Arag{\'o}n-Salamanca}, Merrifield,
  Peterken, Sazonova, Haggar, Bizyaev, Brownstein, Lane, \& Pan}]{greener2022}
Greener, M.~J., {Arag{\'o}n-Salamanca}, A., Merrifield, M., {et~al.} 2022,
  MNRAS, 516, 1275

\bibitem[{Gronke {et~al.}(2022)Gronke, Oh, Ji, \& Norman}]{gronke2022}
Gronke, M., Oh, S.~P., Ji, S., \& Norman, C. 2022, MNRAS, 511, 859

\bibitem[{Guo {et~al.}(2023)Guo, Bacon, Wisotzki, Garel, Blaizot, Schaye,
  Matthee, Leclercq, Boogaard, Richard, Verhamme, Brinchmann, {Michel-Dansac},
  \& Kusakabe}]{guo2023}
Guo, Y., Bacon, R., Wisotzki, L., {et~al.} 2023, Spatially-Resolved
  {{Spectroscopic Analysis}} of {{Ly}}\${\textbackslash}alpha\$ {{Haloes}}:
  {{Radial Evolution}} of the {{Ly}}\${\textbackslash}alpha\$ {{Line Profile}}
  out to 60 Kpc

\bibitem[{Harris {et~al.}(2020)Harris, Millman, {van der Walt}, Gommers,
  Virtanen, Cournapeau, Wieser, Taylor, Berg, Smith, Kern, Picus, Hoyer, {van
  Kerkwijk}, Brett, Haldane, {del R{\'i}o}, Wiebe, Peterson,
  {G{\'e}rard-Marchant}, Sheppard, Reddy, Weckesser, Abbasi, Gohlke, \&
  Oliphant}]{harris2020}
Harris, C.~R., Millman, K.~J., {van der Walt}, S.~J., {et~al.} 2020, Nature,
  585, 357

\bibitem[{Hartigan \& Hartigan(1985)}]{hartigan1985}
Hartigan, J.~A., \& Hartigan, P.~M. 1985, Ann. Statist., 13, 70

\bibitem[{Henry {et~al.}(2021)Henry, Rafelski, Sunnquist, Pirzkal, Pacifici,
  Atek, Bagley, Baronchelli, Barro, Bunker, Colbert, Dai, Elmegreen, Elmegreen,
  Finkelstein, Kocevski, Koekemoer, Malkan, Martin, Mehta, Pahl, Papovich,
  Rutkowski, S{\'a}nchez~Almeida, Scarlata, Snyder, \& Teplitz}]{henry2021}
Henry, A., Rafelski, M., Sunnquist, B., {et~al.} 2021, ApJ, 919, 143

\bibitem[{Ho {et~al.}(2017)Ho, Martin, Kacprzak, \& Churchill}]{ho2017}
Ho, S.~H., Martin, C.~L., Kacprzak, G.~G., \& Churchill, C.~W. 2017, ApJ, 835,
  267

\bibitem[{Hopkins {et~al.}(2021)Hopkins, Chan, Ji, Hummels, Keres, Quataert, \&
  {Faucher-Giguere}}]{hopkins2021}
Hopkins, P.~F., Chan, T.~K., Ji, S., {et~al.} 2021, MNRAS, 501, 3640.
\newblock \doarXiv{2002.02462}

\bibitem[{Huang {et~al.}(2021)Huang, Chen, Shectman, Johnson, Zahedy, Helsby,
  Gauthier, \& Thompson}]{huang2021a}
Huang, Y.-H., Chen, H.-W., Shectman, S.~A., {et~al.} 2021, MNRAS, 502, 4743

\bibitem[{Hunter(2007)}]{hunter2007}
Hunter, J.~D. 2007, Computing in Science and Engineering, 9, 90

\bibitem[{Jansen {et~al.}(2001)Jansen, Lumb, Altieri, Clavel, Ehle, Erd,
  Gabriel, Guainazzi, Gondoin, Much, Munoz, Santos, Schartel, Texier, \&
  Vacanti}]{jansen2001}
Jansen, F., Lumb, D., Altieri, B., {et~al.} 2001, A\&A, 365, L1

\bibitem[{Kacprzak {et~al.}(2012)Kacprzak, Churchill, \&
  Nielsen}]{kacprzak2012}
Kacprzak, G.~G., Churchill, C.~W., \& Nielsen, N.~M. 2012, ApJ, 760, L7.
\newblock \doarXiv{1205.0245}

\bibitem[{Kacprzak {et~al.}(2019)Kacprzak, Pointon, Nielsen, Churchill,
  Muzahid, \& Charlton}]{kacprzak2019}
Kacprzak, G.~G., Pointon, S.~K., Nielsen, N.~M., {et~al.} 2019, ApJ, 886, 91.
\newblock \doarXiv{1910.04310}

\bibitem[{Karki {et~al.}(2023)Karki, Kulkarni, Weng, P{\'e}roux, Augustin,
  Hayes, Ayromlou, Kacprzak, Howk, Szakacs, Klitsch, Hamanowicz, Fresco, Zwaan,
  Biggs, Fox, Kassin, \& Kuntschner}]{karki2023}
Karki, A., Kulkarni, V.~P., Weng, S., {et~al.} 2023, MNRAS, 524, 5524

\bibitem[{Keeney {et~al.}(2013)Keeney, Stocke, Rosenberg, Danforth,
  {Ryan-Weber}, Shull, Savage, \& Green}]{keeney2013}
Keeney, B.~A., Stocke, J.~T., Rosenberg, J.~L., {et~al.} 2013, ApJ, 765, 27

\bibitem[{Kennicutt(1998)}]{kennicutt1998}
Kennicutt, J. 1998, ARA\&A, 36, 189.
\newblock \doarXiv{astro-ph/9807187}

\bibitem[{Kewley {et~al.}(2004)Kewley, Geller, \& Jansen}]{kewley2004}
Kewley, L.~J., Geller, M.~J., \& Jansen, R.~A. 2004, AJ, 127, 2002.
\newblock \doarXiv{astro-ph/0401172}

\bibitem[{Kewley {et~al.}(2019)Kewley, Nicholls, \& Sutherland}]{kewley2019}
Kewley, L.~J., Nicholls, D.~C., \& Sutherland, R.~S. 2019, ARA\&A, 57, 511

\bibitem[{Kluyver {et~al.}(2016)Kluyver, {Ragan-Kelley}, P{\'e}rez, Granger,
  Bussonnier, Frederic, Kelley, Hamrick, Grout, Corlay, Ivanov, Avila, Abdalla,
  Willing, \& {Jupyter Development Team}}]{kluyver2016}
Kluyver, T., {Ragan-Kelley}, B., P{\'e}rez, F., {et~al.} 2016, Jupyter
  {{Notebooks}}---a Publishing Format for Reproducible Computational Workflows,
  87--90

\bibitem[{Koushan {et~al.}(2021)Koushan, Driver, Bellstedt, Davies, Robotham,
  Lagos, Hashemizadeh, Obreschkow, Thorne, Bremer, Holwerda, Hopkins, Jarvis,
  Siudek, \& Windhorst}]{koushan2021}
Koushan, S., Driver, S.~P., Bellstedt, S., {et~al.} 2021, Monthly Notices of
  the Royal Astronomical Society, 503, 2033

\bibitem[{Kuschel {et~al.}(2022)Kuschel, Scarlata, Mehta, Teplitz, Rafelski,
  Wang, Sunnquist, Prichard, Grogin, Windhorst, Rutkowski, Alavi, Chartab,
  Conselice, Dai, Gawiser, Giavalisco, Arrabal~Haro, Hathi, Jansen, Ji,
  Koekemoer, Lucas, Mantha, Mobasher, O'Connell, Robertson, Sattari, Yung,
  Dave, DeMello, Dickinson, Ferguson, Finkelstein, Hayes, Howell, Kaviraj,
  Mackenty, \& Siana}]{kuschel2022}
Kuschel, M., Scarlata, C., Mehta, V., {et~al.} 2022, ApJ, 947, 17

\bibitem[{Lan \& Fukugita(2017)}]{lan2017}
Lan, T.-W., \& Fukugita, M. 2017, ApJ, 850, 156

\bibitem[{Lauger {et~al.}(2005)Lauger, Burgarella, \& Buat}]{lauger2005}
Lauger, S., Burgarella, D., \& Buat, V. 2005, A\&A, 434, 77

\bibitem[{Longobardi {et~al.}(2023)Longobardi, Fossati, Fumagalli, Agarwal,
  Lofthouse, Galbiati, Dutta, Berg, \& Welsh}]{longobardi2023}
Longobardi, A., Fossati, M., Fumagalli, M., {et~al.} 2023, RAS Techniques and
  Instruments, 2, 470

\bibitem[{Lopez {et~al.}(2018)Lopez, Tejos, Ledoux, Barrientos, Sharon, Rigby,
  Gladders, Bayliss, \& Pessa}]{lopez2018}
Lopez, S., Tejos, N., Ledoux, C., {et~al.} 2018, Nature, 554, 493

\bibitem[{Lopez {et~al.}(2020)Lopez, Tejos, Barrientos, Ledoux, Sharon,
  Katsianis, Florian, {Rivera-Thorsen}, Bayliss, Dahle, {Fernandez-Figueroa},
  Gladders, Gronke, Hamel, Pessa, \& Rigby}]{lopez2020}
Lopez, S., Tejos, N., Barrientos, L.~F., {et~al.} 2020, MNRAS, 491, 4442.
\newblock \doarXiv{1911.04809}

\bibitem[{Lundgren {et~al.}(2021)Lundgren, Creech, Brammer, Kirse, Peek, Wake,
  York, Chisholm, Erb, Kulkarni, Straka, Tremonti, \& {van
  Dokkum}}]{lundgren2021}
Lundgren, B.~F., Creech, S., Brammer, G., {et~al.} 2021, ApJ, 913, 50

\bibitem[{Lusso {et~al.}(2019)Lusso, Fumagalli, Fossati, Mackenzie, Bielby,
  Arrigoni~Battaia, Cantalupo, Cooke, Cristiani, Dayal, D'Odorico, Haardt,
  Lofthouse, Morris, Peroux, Prichard, Rafelski, Simcoe, Swinbank, \&
  Theuns}]{lusso2019}
Lusso, E., Fumagalli, M., Fossati, M., {et~al.} 2019, MNRAS, 485, L62

\bibitem[{Lusso {et~al.}(2023)Lusso, Nardini, Fumagalli, Fossati,
  Arrigoni~Battaia, Revalski, Rafelski, D'Odorico, Peroux, Cristiani, Dayal,
  Haardt, \& Lofthouse}]{lusso2023}
Lusso, E., Nardini, E., Fumagalli, M., {et~al.} 2023, MNRAS, 525, 4388

\bibitem[{Madau \& Dickinson(2014)}]{madau2014}
Madau, P., \& Dickinson, M. 2014, ARA\&A, 52, 415.
\newblock \doarXiv{1403.0007}

\bibitem[{Martin {et~al.}(2019)Martin, Ho, Kacprzak, \& Churchill}]{martin2019}
Martin, C.~L., Ho, S.~H., Kacprzak, G.~G., \& Churchill, C.~W. 2019, ApJ, 878,
  84

\bibitem[{Massey~Jr.(1951)}]{masseyjr.1951}
Massey~Jr., F.~J. 1951, Journal of the American Statistical Association, 46, 68

\bibitem[{Mitchell {et~al.}(2020)Mitchell, Schaye, Bower, \&
  Crain}]{mitchell2020}
Mitchell, P.~D., Schaye, J., Bower, R.~G., \& Crain, R.~A. 2020, MNRAS, 494,
  3971

\bibitem[{Mortensen {et~al.}(2021)Mortensen, Keerthi~Vasan, Jones,
  {Faucher-Gigu{\`e}re}, Sanders, Ellis, Leethochawalit, \&
  Stark}]{mortensen2021}
Mortensen, K., Keerthi~Vasan, G.~C., Jones, T., {et~al.} 2021, ApJ, 914, 92

\bibitem[{Nedkova {et~al.}(2021)Nedkova, H{\"a}u{\ss}ler, Marchesini, Dimauro,
  Brammer, Eigenthaler, Feinstein, Ferguson, {Huertas-Company}, Johnston,
  {Kado-Fong}, Kartaltepe, Labb{\'e}, {Lange-Vagle}, Martis, McGrath, Muzzin,
  Oesch, {Ordenes-Brice{\~n}o}, Puzia, Shipley, Simmons, Skelton, Stefanon,
  {van der Wel}, \& Whitaker}]{nedkova2021}
Nedkova, K.~V., H{\"a}u{\ss}ler, B., Marchesini, D., {et~al.} 2021, MNRAS, 506,
  928

\bibitem[{Nelson {et~al.}(2019)Nelson, Pillepich, Springel, Pakmor, Weinberger,
  Genel, Torrey, Vogelsberger, Marinacci, \& Hernquist}]{nelson2019}
Nelson, D., Pillepich, A., Springel, V., {et~al.} 2019, MNRAS, 490, 3234.
\newblock \doarXiv{1902.05554}

\bibitem[{Ng {et~al.}(2019)Ng, Nielsen, Kacprzak, Pointon, Muzahid, Churchill,
  \& Charlton}]{ng2019}
Ng, M., Nielsen, N.~M., Kacprzak, G.~G., {et~al.} 2019, ApJ, 886, 66

\bibitem[{Nielsen {et~al.}(2018)Nielsen, Kacprzak, Pointon, Churchill, \&
  Murphy}]{nielsen2018}
Nielsen, N.~M., Kacprzak, G.~G., Pointon, S.~K., Churchill, C.~W., \& Murphy,
  M.~T. 2018, ApJ, 869, 153

\bibitem[{Pandya {et~al.}(2021)Pandya, Fielding, {Angl{\'e}s-Alc{\'a}zar},
  Somerville, Bryan, Hayward, Stern, Kim, Quataert, Forbes,
  {Faucher-Gigu{\`e}re}, Feldmann, Hafen, Hopkins, Kere{\v s}, Murray, \&
  Wetzel}]{pandya2021}
Pandya, V., Fielding, D.~B., {Angl{\'e}s-Alc{\'a}zar}, D., {et~al.} 2021,
  MNRAS, 508, 2979

\bibitem[{Peng {et~al.}(2002)Peng, Ho, Impey, \& Rix}]{peng2002}
Peng, C.~Y., Ho, L.~C., Impey, C.~D., \& Rix, H.-W. 2002, AJ, 124, 266.
\newblock \doarXiv{astro-ph/0204182}

\bibitem[{Peng {et~al.}(2010)Peng, Ho, Impey, \& Rix}]{peng2010a}
---. 2010, AJ, 139, 2097

\bibitem[{Perez \& Granger(2007)}]{perez2007}
Perez, F., \& Granger, B.~E. 2007, Computing in Science and Engineering, 9, 21

\bibitem[{P{\'e}roux {et~al.}(2020)P{\'e}roux, Nelson, {van de Voort},
  Pillepich, Marinacci, Vogelsberger, \& Hernquist}]{peroux2020a}
P{\'e}roux, C., Nelson, D., {van de Voort}, F., {et~al.} 2020, MNRAS, 499, 2462

\bibitem[{Pirard {et~al.}(2004)Pirard, {Kissler-Patig}, Moorwood, Biereichel,
  Delabre, Dorn, Finger, Gojak, Huster, Jung, Koch, Le~Louarn, Lizon, Mehrgan,
  Pozna, Silber, Sokar, \& Stegmeier}]{pirard2004}
Pirard, J.-F., {Kissler-Patig}, M., Moorwood, A., {et~al.} 2004, Ground-based
  Instrumentation for Astronomy, 5492, 1763

\bibitem[{{Planck Collaboration}(2020)}]{planckcollaboration2020}
{Planck Collaboration}. 2020, A\&A, 641, A6

\bibitem[{Pointon {et~al.}(2019)Pointon, Kacprzak, Nielsen, Muzahid, Murphy,
  Churchill, \& Charlton}]{pointon2019}
Pointon, S.~K., Kacprzak, G.~G., Nielsen, N.~M., {et~al.} 2019, ApJ, 883, 78.
\newblock \doarXiv{1907.05557}

\bibitem[{Prochaska {et~al.}(2019)Prochaska, Burchett, Tripp, Werk, Willmer,
  Howk, Lange, Tejos, Meiring, Tumlinson, Lehner, Ford, \&
  Dave}]{prochaska2019}
Prochaska, J.~X., Burchett, J.~N., Tripp, T.~M., {et~al.} 2019, ApJS, 243, 24.
\newblock \doarXiv{1908.07675}

\bibitem[{Rafelski {et~al.}(2015)Rafelski, Teplitz, Gardner, Coe, Bond,
  Koekemoer, Grogin, Kurczynski, McGrath, Bourque, Atek, Brown, Colbert,
  Codoreanu, Ferguson, Finkelstein, Gawiser, Giavalisco, Gronwall, Hanish, Lee,
  Mehta, {de Mello}, Ravindranath, Ryan, Scarlata, Siana, Soto, \&
  Voyer}]{rafelski2015}
Rafelski, M., Teplitz, H.~I., Gardner, J.~P., {et~al.} 2015, The Astronomical
  Journal, 150, 31

\bibitem[{Ramesh \& Nelson(2024)}]{ramesh2024}
Ramesh, R., \& Nelson, D. 2024, Monthly Notices of the Royal Astronomical
  Society, 528, 3320

\bibitem[{Rao {et~al.}(2011)Rao, {Belfort-Mihalyi}, Turnshek, Monier, Nestor,
  \& Quider}]{rao2011}
Rao, S.~M., {Belfort-Mihalyi}, M., Turnshek, D.~A., {et~al.} 2011, MNRAS, 416,
  1215

\bibitem[{Revalski {et~al.}(2023)Revalski, Rafelski, Fumagalli, Fossati,
  Pirzkal, Sunnquist, Prichard, Henry, Bagley, Dutta, Papini, Battaia,
  D'Odorico, Dayal, {Estrada-Carpenter}, Lofthouse, Lusso, Morris, Nedkova,
  Papovich, \& Peroux}]{revalski2023}
Revalski, M., Rafelski, M., Fumagalli, M., {et~al.} 2023, ApJS, 265, 40

\bibitem[{Revalski {et~al.}(2024)Revalski, Rafelski, Henry, Fossati, Fumagalli,
  Dutta, Pirzkal, Beckett, Arrigoni~Battaia, Dayal, D'Odorico, Lusso, Nedkova,
  Prichard, Papovich, \& Peroux}]{revalski2024}
Revalski, M., Rafelski, M., Henry, A., {et~al.} 2024, The {{MUSE Ultra Deep
  Field}} ({{MUDF}}). {{V}}. {{Characterizing}} the {{Mass-Metallicity
  Relation}} for {{Low Mass Galaxies}} at \${\textbackslash}boldsymbol\{z\}\$
  \${\textbackslash}boldsymbol\{{\textbackslash}sim\}\$ 1 - 2

\bibitem[{{Roberts-Borsani}(2020)}]{roberts-borsani2020}
{Roberts-Borsani}, G.~W. 2020, MNRAS, 494, 4266

\bibitem[{{Rodriguez-Gomez} {et~al.}(2019){Rodriguez-Gomez}, Snyder, Lotz,
  Nelson, Pillepich, Springel, Genel, Weinberger, Tacchella, Pakmor, Torrey,
  Marinacci, Vogelsberger, Hernquist, \& Thilker}]{rodriguez-gomez2019}
{Rodriguez-Gomez}, V., Snyder, G.~F., Lotz, J.~M., {et~al.} 2019, MNRAS, 483,
  4140

\bibitem[{Rodr{\'i}guez~Montero {et~al.}(2023)Rodr{\'i}guez~Montero,
  {Martin-Alvarez}, Slyz, Devriendt, Dubois, \& Sijacki}]{rodriguezmontero2023}
Rodr{\'i}guez~Montero, F., {Martin-Alvarez}, S., Slyz, A., {et~al.} 2023, The
  Impact of Cosmic Rays on the Interstellar Medium and Galactic Outflows of
  {{Milky Way}} Analogues

\bibitem[{Rubin {et~al.}(2012)Rubin, Prochaska, Koo, \& Phillips}]{rubin2012}
Rubin, K. H.~R., Prochaska, J.~X., Koo, D.~C., \& Phillips, A.~C. 2012, ApJL,
  747, L26

\bibitem[{Rubin {et~al.}(2014)Rubin, Prochaska, Koo, Phillips, Martin, \&
  Winstrom}]{rubin2014}
Rubin, K. H.~R., Prochaska, J.~X., Koo, D.~C., {et~al.} 2014, ApJ, 794, 156

\bibitem[{Sanders {et~al.}(2018)Sanders, Shapley, Kriek, Freeman, Reddy, Siana,
  Coil, Mobasher, Dav{\'e}, Shivaei, Azadi, Price, Leung, Fetherholf, de~Groot,
  Zick, Fornasini, \& Barro}]{sanders2018}
Sanders, R.~L., Shapley, A.~E., Kriek, M., {et~al.} 2018, ApJ, 858, 99

\bibitem[{Sanders {et~al.}(2021)Sanders, Shapley, Jones, Reddy, Kriek, Siana,
  Coil, Mobasher, Shivaei, Dav{\'e}, Azadi, Price, Leung, Freeman, Fetherolf,
  de~Groot, Zick, \& Barro}]{sanders2021}
Sanders, R.~L., Shapley, A.~E., Jones, T., {et~al.} 2021, ApJ, 914, 19

\bibitem[{Schroetter {et~al.}(2019)Schroetter, Bouch{\'e}, Zabl, Contini,
  Wendt, Schaye, Mitchell, Muzahid, Marino, Bacon, Lilly, Richard, \&
  Wisotzki}]{schroetter2019}
Schroetter, I., Bouch{\'e}, N.~F., Zabl, J., {et~al.} 2019, MNRAS, 490, 4368.
\newblock \doarXiv{1907.09967}

\bibitem[{Silverman(1981)}]{silverman1981}
Silverman, B.~W. 1981, Journal of the Royal Statistical Society: Series B
  (Methodological), 43, 97

\bibitem[{Somerville \& Dav{\'e}(2015)}]{somerville2015}
Somerville, R.~S., \& Dav{\'e}, R. 2015, ARA\&A, 53, 51

\bibitem[{Sorgho {et~al.}(2019)Sorgho, Foster, Carignan, \&
  Chemin}]{sorgho2019}
Sorgho, A., Foster, T., Carignan, C., \& Chemin, L. 2019, MNRAS.
\newblock \doarXiv{1903.03767}

\bibitem[{Steidel {et~al.}(2010)Steidel, Erb, Shapley, Pettini, Reddy,
  Bogosavljevi{\'c}, Rudie, \& Rakic}]{steidel2010}
Steidel, C.~C., Erb, D.~K., Shapley, A.~E., {et~al.} 2010, ApJ, 717, 289

\bibitem[{Stern {et~al.}(2024)Stern, Fielding, Hafen, Su, Naor,
  {Faucher-Gigu{\`e}re}, Quataert, \& Bullock}]{stern2024}
Stern, J., Fielding, D., Hafen, Z., {et~al.} 2024, MNRAS, 530, 1711

\bibitem[{Tortora {et~al.}(2022)Tortora, Hunt, \& Ginolfi}]{tortora2022}
Tortora, C., Hunt, L.~K., \& Ginolfi, M. 2022, A\&A, 657, A19

\bibitem[{Tumlinson {et~al.}(2017)Tumlinson, Peeples, \& Werk}]{tumlinson2017}
Tumlinson, J., Peeples, M.~S., \& Werk, J.~K. 2017, ARA\&A, 55, 389.
\newblock \doarXiv{1709.09180}

\bibitem[{Verhamme {et~al.}(2018)Verhamme, Garel, Ventou, Contini, Bouch{\'e},
  Herenz, Richard, Bacon, Schmidt, Maseda, Marino, Brinchmann, Cantalupo,
  Caruana, Cl{\'e}ment, Diener, Drake, Hashimoto, Inami, Kerutt, Kollatschny,
  Leclercq, Patr{\'i}cio, Schaye, Wisotzki, \& Zabl}]{verhamme2018}
Verhamme, A., Garel, T., Ventou, E., {et~al.} 2018, MNRAS, 478, L60

\bibitem[{Virtanen {et~al.}(2020)Virtanen, Gommers, Oliphant, Haberland, Reddy,
  Cournapeau, Burovski, Peterson, Weckesser, Bright, {van der Walt}, Brett,
  Wilson, Millman, Mayorov, Nelson, Jones, Kern, Larson, Carey, Polat, Feng,
  Moore, VanderPlas, Laxalde, Perktold, Cimrman, Henriksen, Quintero, Harris,
  Archibald, Ribeiro, Pedregosa, {van Mulbregt}, \& {SciPy 1. 0
  Contributors}}]{virtanen2020}
Virtanen, P., Gommers, R., Oliphant, T.~E., {et~al.} 2020, Nature Methods, 17,
  261

\bibitem[{Vulcani {et~al.}(2014)Vulcani, Bamford, H{\"a}u{\ss}ler, Vika, Rojas,
  Agius, Baldry, Bauer, Brown, Driver, Graham, Kelvin, Liske, Loveday, Popescu,
  Robotham, \& Tuffs}]{vulcani2014}
Vulcani, B., Bamford, S.~P., H{\"a}u{\ss}ler, B., {et~al.} 2014, MNRAS, 441,
  1340

\bibitem[{Wendt {et~al.}(2021)Wendt, Bouch{\'e}, Zabl, Schroetter, \&
  Muzahid}]{wendt2021}
Wendt, M., Bouch{\'e}, N.~F., Zabl, J., Schroetter, I., \& Muzahid, S. 2021,
  MNRAS, 502, 3733

\bibitem[{Weng {et~al.}(2023)Weng, P{\'e}roux, Karki, Augustin, Kulkarni,
  Hamanowicz, Zwaan, Sadler, Nelson, Hayes, Kacprzak, Fox, Bollo, Casavecchia,
  \& Szakacs}]{weng2023a}
Weng, S., P{\'e}roux, C., Karki, A., {et~al.} 2023, MNRAS, 523, 676

\bibitem[{Wetzel {et~al.}(2013)Wetzel, Tinker, Conroy, \& {van den
  Bosch}}]{wetzel2013}
Wetzel, A.~R., Tinker, J.~L., Conroy, C., \& {van den Bosch}, F.~C. 2013,
  MNRAS, 432, 336

\bibitem[{Wild {et~al.}(2008)Wild, Kauffmann, White, York, Lehnert, Heckman,
  Hall, Khare, Lundgren, Schneider, \& Berk}]{wild2008}
Wild, V., Kauffmann, G., White, S., {et~al.} 2008, MNRAS, 388, 227.
\newblock \doarXiv{0802.4100}

\bibitem[{Wilde {et~al.}(2021)Wilde, Werk, Burchett, Prochaska, Tchernyshyov,
  Tripp, Tejos, Lehner, Bordoloi, O'Meara, \& Tumlinson}]{wilde2021}
Wilde, M.~C., Werk, J.~K., Burchett, J.~N., {et~al.} 2021, ApJ, 912, 9

\bibitem[{Wootten \& Thompson(2009)}]{wootten2009}
Wootten, A., \& Thompson, A.~R. 2009, IEEE Proceedings, 97, 1463

\bibitem[{Xu {et~al.}(2023)Xu, Heckman, Henry, Berg, Chisholm, James, Martin,
  Stark, Hayes, {Arellano-C{\'o}rdova}, Carr, Huberty, Mingozzi, Scarlata, \&
  Sugahara}]{xu2023}
Xu, X., Heckman, T., Henry, A., {et~al.} 2023, ApJ, 948, 28

\bibitem[{Yang {et~al.}(2022)Yang, Scholte, \& Saintonge}]{yang2022}
Yang, N., Scholte, D., \& Saintonge, A. 2022, MNRAS

\bibitem[{Zabl {et~al.}(2021)Zabl, Bouch{\'e}, Wisotzki, Schaye, Leclercq,
  Garel, Wendt, Schroetter, Muzahid, Cantalupo, Contini, Bacon, Brinchmann, \&
  Richard}]{zabl2021}
Zabl, J., Bouch{\'e}, N.~F., Wisotzki, L., {et~al.} 2021, MNRAS, 507, 4294

\bibitem[{Zhang {et~al.}(2023)Zhang, Cai, Xu, Afruni, Wu, Wang, Battaia, Li,
  Wang, \& Bi}]{zhang2023}
Zhang, S., Cai, Z., Xu, D., {et~al.} 2023, ApJ, 952, 124

\end{thebibliography}
\bibliographystyle{aasjournal}



\end{document}